\documentclass{article}
\usepackage{amssymb}
\usepackage{cite,multirow} 
\usepackage[dvips]{graphicx,epsfig}
\usepackage{graphics}
\def\1ad{\mbox{\normalsize $^1$}}
\def\2ad{\mbox{\normalsize $^2$}}
\def\3ad{\mbox{\normalsize $^3$}}
\def\4ad{\mbox{\normalsize $^4$}}
\def\5ad{\mbox{\normalsize $^5$}}
\def\6ad{\mbox{\normalsize $^6$}}
\def\7ad{\mbox{\normalsize $^7$}}
\def\8ad{\mbox{\normalsize $^8$}}

\makeatletter
\@addtoreset{equation}{section}
\makeatother
\renewcommand{\theequation}{\thesection.\arabic{equation}}
 
\def\beq{\begin{equation}}                     %  
\def\eeq{\end{equation}}                       % 
\def\bea{\begin{eqnarray}}                     %         % 
\def\eea{\end{eqnarray}}                       %       %  
\def\0 {\nonumber}

\begin{document}

\setcounter{page}{0}
\begin{titlepage}
\titlepage
\rightline{hep-th/0506232}
\rightline{Bicocca-FT-05-16}
\vskip 3cm
\centerline{{ \bf \Large R-charges from toric diagrams and  }}
\vskip 0.5cm
\centerline{{ \bf \Large the equivalence of a-maximization and Z-minimization}}
\vskip 1cm
\centerline{{\bf Agostino Butti 
 and Alberto Zaffaroni}}
\vskip 1.5truecm
\begin{center}
\em 
$^a$ Dipartimento di Fisica, Universit\`{a} di Milano-Bicocca \\ 
P.zza della Scienza, 3; I-20126 Milano, Italy\\
\vskip .4cm

\vskip 2.5cm
\end{center}
\begin{abstract}
We conjecture a general formula for assigning R-charges and multiplicities
for the chiral fields of all gauge theories living on branes at toric
singularities. We check that the central charge and the dimensions of all the
chiral fields agree with the information on volumes that can be extracted
from toric geometry. We also analytically check the equivalence between
the volume minimization procedure discovered in hep-th/0503183 and 
a-maximization, for the most general toric diagram. 
Our results can be considered as a very general check
of the AdS/CFT correspondence, valid for all superconformal theories
associated with toric singularities.

\vskip1cm

\end{abstract}
\vskip 0.5\baselineskip

\vfill
 \hrule width 5.cm
\vskip 2.mm
{\small
\noindent agostino.butti@mib.infn.it\\
alberto.zaffaroni@mib.infn.it}
\begin{flushleft}
%\vspace{.5cm}
\end{flushleft}
\end{titlepage}
\large
\section{Introduction}

D3 branes living at conical Calabi-Yau 
singularities are a good laboratory for the AdS/CFT correspondence since 
its early days. The world-volume
theory on the branes is dual to a type  IIB background
of the form $AdS_5\times H$, where $H$ is the horizon manifold 
\cite{kw,horizon}. Supersymmetry requires that $H$ is a Sasaki-Einstein 
manifold. Until few months ago, the only known Sasaki-Einstein metrics
were the round sphere $S^5$ and $T^{1,1}$, the horizon of the conifold.
Recently, various infinite classes of new regular Sasaki-Einstein metrics 
were constructed \cite{gauntlett,CLPP,MSL} and named $Y^{p,q}$
and $L^{p,q,r}$. For infinite values of the integers $p,q,r$ one obtains
smooth Sasaki-Einstein manifolds.
With the determination of the corresponding dual gauge theory 
(see \cite{benvenuti} for the $Y^{p,q}$ manifolds and \cite{kru2,noi,tomorrow}
for the $L^{p,q,r}$),
new checks of the AdS/CFT correspondence were possible \cite{bertolini,benvenuti,kleb,kru,kru2,noi,tomorrow}. As well known,
the central charge of the CFT and the dimension of some operators can be
compared with the volumes of $H$ and of some of its submanifolds.
In particular, the a-maximization technique \cite{intriligator} now
allows for a detailed computation of the relevant quantum field theory
quantities. Needless to say, the agreement of the two computations
is perfect. 

The number of explicit metrics for Sasaki-Einstein horizons than can be used
in the AdS/CFT correspondence is rapidly increasing. 
However, to demystify a little bit the importance of having an
explicit metric, we should note that all relevant volumes are computed
for calibrated divisors. This means that these volumes can be computed
without actually knowing the metric. There exist moreover a beautiful geometrical counterpart of the a-maximization \cite{intriligator}: this is the
volume minimization proposed in \cite{MSY} for determining the Reeb vector
for toric cones. This procedure only relies on the vectors defining the 
toric fan. 
This suggests that with a correspondence between toric diagrams and gauge theories, many checks of the AdS/CFT correspondence can be done without an explicit knowledge of the metric. It is the purpose of this paper, indeed, to show
that the knowledge of the toric data is sufficient to determine 
many properties of the dual gauge theory and to perform all the
mentioned checks, for every singularity.

The precise correspondence between conical Calabi-Yau singularities and superconformal gauge theories is still unknown. However,
a remarkable progress has been recently made for the class of
Gorenstein toric singularities. The brane tiling (dimers) construction
\cite{dimers}, an ingenious generalization of the Brane Boxes \cite{boxes,boxes2},
introduces a direct relation between an Hanany-Witten realization \cite{hw}
for gauge theory and the toric diagram. In particular, from the quiver
associated with a non-chiral superconformal gauge theory one
can determine the dual brane tiling configuration, a dimer lattice.
It is then possible to associate a toric diagram with each of these lattices, 
identifying the dual Calabi-Yau. The inverse process
(to associate a gauge theory with a given singularity) is more difficult. 
However, for the mentioned checks of the AdS/CFT correspondence, we don't
really need the full quiver description of the gauge theory.
We just need to know the R-charges and the multiplicities of 
chiral fields. In this paper, elaborating on existing results in the literature \cite{hananymirror,kru2,benvenuti,tomorrow}, 
we propose a general assignment of charges and multiplicities for the gauge
theory dual to a generic Gorenstein singularity. This assignment is made
using only the toric data of the singularity.
We then compare the result of a-maximization with that of volume minimization 
showing that the two procedures are completely equivalent. 
This agreement is remarkable.
We have two different algebraic procedures for computing
the R-symmetry charges of the fields and the volumes. The first is based
on the maximization of the central charge \cite{intriligator}. 
The second one can be
efficiently encoded in a geometrical minimization procedure for determining
the Reeb vector \cite{MSY}. The two procedures deal with
different test quantities (the R-charges on one side and the components
of the Reeb vector on the other) and with different functions to be extremized.
However, we will show that, with a suitable parametrization, 
the two functions ($a$ and the inverse volume) are equal, even before
extremization.

The agreement of results in the gauge theory and the supergravity side can be 
regarded as a general non-trivial check of the AdS/CFT 
correspondence, valid for all the theories living on branes at
toric singularities.  

The paper is organized as follows. In Section \ref{gauge} 
we briefly review the general features of the gauge theories dual to
conical singularities. In Section \ref{geometry} we propose the 
assignment of R-charges and multiplicities for the gauge theory in terms
of geometrical data. In Section \ref{comparison} we show the equivalence
of the a-maximization and the volume minimization. %, sketching an analytic proof. 
Section \ref{examples}
contains several examples based on known gauge theories and various
observations. In particular, as a by product of our analysis, we discuss
in detail the case of the manifolds $X^{p,q}$ introduced in \cite{hananyX}
whose general analysis was missing in the literature. We also make
some observations on the identification of fields using the brane tiling
technology. Finally, the Appendix contains the proofs of various
results that are too long and boring for the main text. 

\section{Generalities about the gauge theory}
\label{gauge}
We consider $N$ D3-branes living at a conical Gorenstein singularity.
The internal manifold is a six-dimensional symplectic toric cone; its
base, or horizon, is a five-dimensional Sasaki-Einstein manifold $H$ 
\cite{kw,horizon}. As well known, the ${\cal N}=1$ gauge theory living on the branes is 
superconformal and dual in the AdS/CFT correspondence to the type IIB
background $AdS_5\times H$. 
The gauge theory on the world-volume of the D3 branes is not chiral
and represents a toric phase \cite{bo}, where 
all gauge groups have the same number of colors $N$ and the only matter
fields are bi-fundamentals. By applying a Seiberg duality we can obtain
a different theory that flows in the IR to the same CFT. If we dualize
a gauge group with number of flavors equal to $2N$ we remain in a toric phase
where all gauge groups have number of colors $N$.
In this process the number of gauge groups remains constant but the
number of matter fields changes.  In a toric phase the following relation between
the number of gauge groups $F$, the number of chiral fields $E$ and the number of 
terms in the superpotential $V$
\begin{equation}
V-E+F=0
\end{equation} 
is valid \cite{dimers}.
Indeed for a gauge theory living on branes placed at the tip of toric CY cone, one can extend the quiver diagram, drawing it on a torus $ T^2$. The dual graph, known as the brane tiling
associated with the gauge theory \cite{dimers}, has $F$ faces, $E$ edges
and $V$ vertices and it is still defined on a torus. The previous formula then follows
from the Euler formula for a torus \cite{dimers}.
 
We can assign an R-charge to all the chiral fields.
The most general non-anomalous R-symmetry is determined by the cancellation of anomalies 
for each gauge group and by the requirement that each term in the superpotential has R-charge
$2$. This would seem to imply $F+V=E$ linear conditions for $E$ unknowns with an unique
solution. However, in the cases we are interested in, not all the conditions 
are linearly independent. This is reflected by the fact that
the R-symmetry can mix with all the non anomalous $U(1)$ global symmetries.
We can count the number of global non-anomalous $U(1)$ symmetries from the 
number of massless vectors in the $AdS$ dual. Since the manifold is toric, 
the metric has three $U(1)$ isometries.
One of these (the Reeb one) corresponds to the R-symmetry while the other two
are related to non-anomalous global $U(1)$s. 
Other gauge fields in $AdS$ come
from the reduction of the RR four form on the non-trivial three-spheres
in the horizon manifold $H$. The number of three-cycles depends on the topology
of the horizon, and, as we will review soon, can be computed using the toric data of the
singularity.
In the supergravity literature the 
vector multiplets obtained from RR four form are known as the Betti multiplets. 
On the gauge theory side,
these gauge fields  correspond to baryonic symmetries.

At the fixed point, 
only one of the possible non-anomalous R-symmetry enters
in the superconformal algebra. It is the one in the same multiplet as the 
stress-energy tensor. The actual value of the R-charges at the fixed point can 
be found by using the a-maximization technique \cite{intriligator}. As shown in
\cite{intriligator}, we have to maximize the a-charge \cite{anselmi}
\begin{equation}
a(R)=\frac{3}{32}(3 {\rm Tr} R^3-{\rm Tr} R)
\end{equation}
It is not difficult to show that the absence of anomalies implies ${\rm Tr} R=0$
so that we can equivalently maximize ${\rm Tr} R^3$.

The results of the maximization give a complete information about 
the values of the central charge and the dimensions of chiral operators
at the conformal fixed point. These can be compared with the prediction
of the AdS/CFT correspondence \cite{gubser,gubserkleb}. 
The first important point is that the central charge is related to the volume 
of the internal manifold 
\cite{gubser}
\begin{equation}
a=\frac{\pi^3}{4 {\rm Vol}(H)}
\label{central}
\end{equation}
Moreover, recall that in the AdS/CFT correspondence a special role is played
by baryons. The gravity dual describes a theory with $SU(N)$ gauge groups. The fact that the groups are $SU(N)$ and
not $U(N)$ allows the existence of dibaryons. Each bi-fundamental field
$\Phi_\alpha^\beta$ gives rise to a gauge invariant baryonic operator
$$ \epsilon^{\alpha_1...\alpha_N} \Phi_{\alpha_1}^{\beta_1}...\Phi_{\alpha_N}^{\beta_N}
\epsilon_{\beta_1...\beta_N}$$
It is sometime convenient
to think about the baryonic symmetries as non-anomalous combinations of $U(1)$ 
factors in the enlarged $\prod  U(N)$ theory.
In the AdS dual the baryonic symmetries correspond to the reduction of the RR four form
and the dibaryons are described by a D3-brane wrapped on a non-trivial three cycle.
The R-charge of the $i$-th field can be computed in terms of
the volume of the corresponding cycle $\Sigma_i$ using the formula \cite{gubserkleb}
\begin{equation}
R_i=\frac{\pi {\rm Vol}(\Sigma_i)}{3 {\rm Vol}(H)}
\label{baryons}
\end{equation}

\section{Geometrical formulae for the R-charges}
\label{geometry}
In this Section we propose a general formula for the R-charges and the multiplicity
of chiral fields based only on the toric data \footnote{For the necessary
elements of toric geometry see \cite{fulton} and the review part of \cite{MS}.}. This proposal is the natural combination of existing results \cite{hananymirror,kru2,benvenuti,tomorrow} and it is substantially implicit in previous
papers on the subject. It is based on a formula for multiplicities first
derived using mirror symmetry \cite{hananymirror}. The same proposal
appeared for the case of $L^{p,q:r,s}$ manifolds in \cite{kru2}, under the name
of ``folded quiver''.

\begin{figure}
\begin{minipage}[t]{0.48\linewidth}
\centering
\includegraphics[scale=0.67]{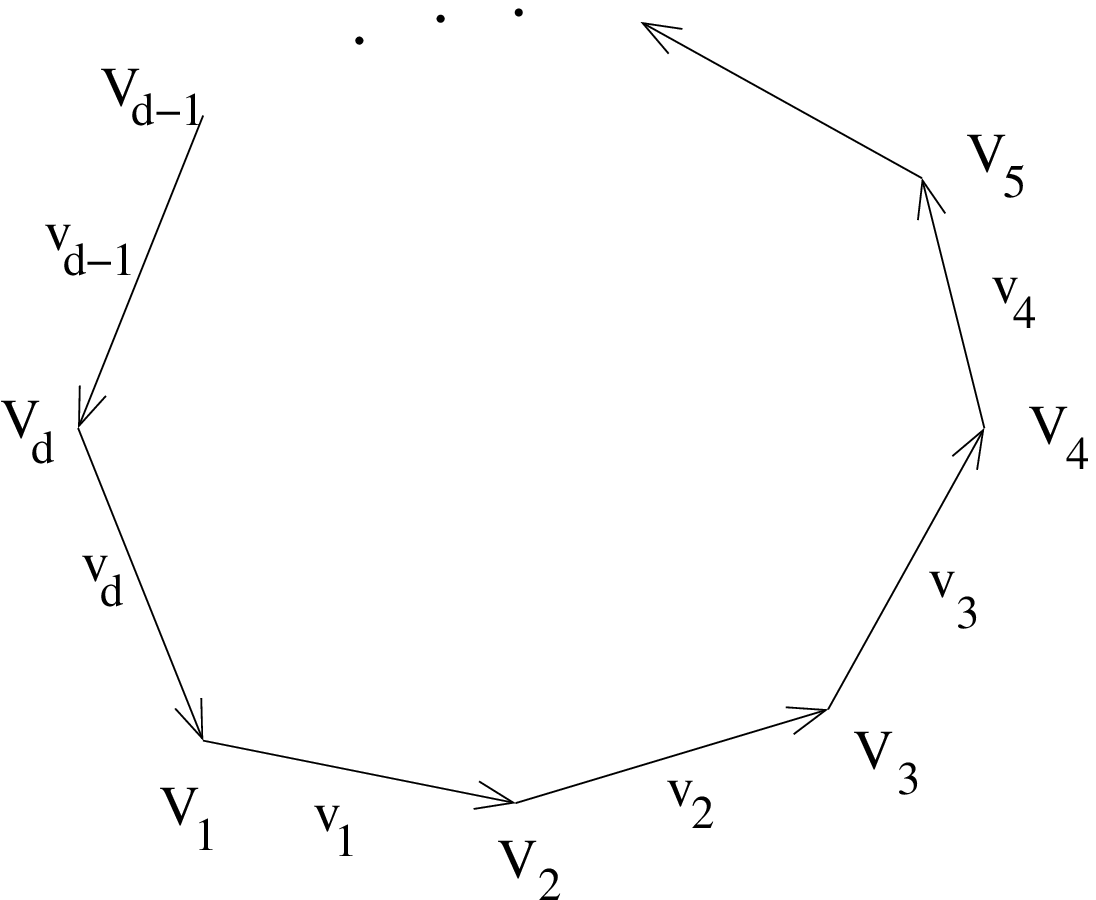}
\caption{The convex polygon $P$.}
\label{polygon}
\end{minipage}%
~~~~~~\begin{minipage}[t]{0.48\linewidth}
\centering
\includegraphics[scale=0.67]{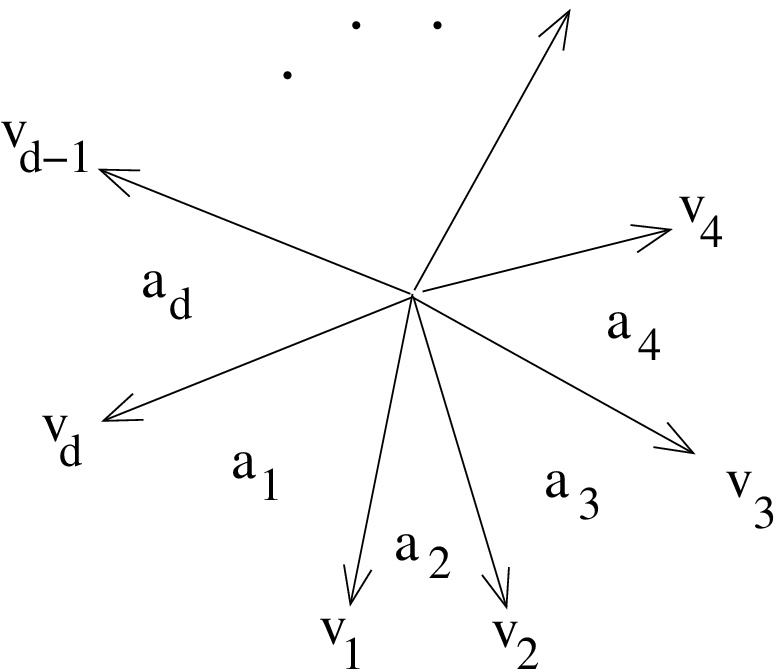}
\caption{The $(p,q)$ web for $P$.}
\label{pqweb}
\end{minipage}%
\end{figure}

The fan associated with a six-dimensional symplectic toric cone is
generated by $d$ integers primitive vectors in $\mathbb R^3$, 
which we call $V_i$, $i=1,2 \ldots d$. 
When the cone is a Calabi-Yau manifold, 
we can perform an $SL(3,\mathbb Z)$ transformation to put the first coordinates of the $V_i$'s equal to 1. 
The intersection of the fan with the plane of points having the first coordinate $x=1$ is thus a convex polygon $P$, called toric diagram, 
and we shall call the vectors associated to its sides $v_i$, 
$i=1,2 \ldots d$, as in Figure \ref{polygon}. 
In Figure \ref{pqweb} we draw the corresponding $(p,q)$ web: the vectors $v_i$ have the same length than the edges of the polygon $P$ \footnote{With a little abuse of notation we call $v_i$ both the sides of $P$ and the vectors of the $(p,q)$ web. In fact they differ only by a rotation of $90^o$. 
When some of the sides of the polygon $P$ pass through integer points, that is for singular horizons, we should consider more complicated $(p,q)$ webs; here we are ignoring such subtleties. We claim that this does not affect the process of a-maximization, since it is equivalent to setting to zero the charges $b_i$ associated with integers points on the sides of $P$ (see subsection \ref{singhorizon}).}. 
Let us also define the symbol:
\begin{equation}
\langle w_i, w_j \rangle \equiv \det(w_i, w_j) 
\end{equation} 
that is the determinant of the matrix with $w_i$ and $w_j$ as first and second line respectively, where $w_i$ and $w_j$ are two vectors in the plane of $P$. This is the oriented area of the parallelogram generated by $w_i$ and $w_j$.

Some of the data of the gauge theory can be extracted directly from the
geometry of the cone. In particular, there exist simple formulae for
the number of gauge groups $F$ and the total 
number of chiral bi-fundamental fields $E$ \cite{hananymirror,benvenuti}
\begin{eqnarray}
F&=& 2{\rm Area}(P)\nonumber\\
E&=& \frac{1}{2}\sum_{i,j} |\langle v_i,v_j\rangle |
\label{numbers}
\end{eqnarray} 
Notice that the expression for $E$ refers to a particular toric phase 
of the gauge theory.  The number of toric phases of a theory
can be large; hopefully, the value of $E$ in formula~(\ref{numbers}) refers
to the phase with the minimal number of fields.
 
The R-charges and the multiplicity of fields with given R-charge are more difficult to determine.
Here we make a proposal based on the following general observations. 
Each chiral field
is associated with a dibaryon and, consequently, with a 
supersymmetric three cycle in the horizon $H$.
The cone over this cycle is a divisor in the symplectic cone $C(H)$.
Each edge $V_i$ in the fan determines a divisor $D_i$ and the collection of the $D_i$,
subject to some relations, 
is a complete basis of divisors for $C(H)$ \cite{fulton}. 
We can therefore associate a type of 
chiral field to each vector $V_i$ \cite{tomorrow} and assign it a trial
R-charge $a_i$. It is important to stress that more than one chiral field
is associated with a single divisor: as pointed out in \cite{tomorrow}, a D3 brane wrapped on the cycle
$\Sigma$ may have more than one supersymmetric vacuum and each of these
corresponds to a different bi-fundamental but with the same R-charge. 
As shown in \cite{MSY} the volumes of the base three cycles $\Sigma_i$ 
of the divisors $D_i$ satisfy the relation
\begin{equation}
\sum_{i=1}^d {\rm Vol}(\Sigma_i)=\frac{6}{\pi} {\rm Vol} (H)
\label{norm1} 
\end{equation}
which implies, using formula ~(\ref{baryons}), $\sum_{i=1}^d a_i=2$.
In general, these $d$ fields will not exhaust all the different types
of chiral fields. We expect the existence of other dibaryons
obtained from divisors which are linear combinations of the $D_i$s.
The R-charges of the corresponding fields will not be independent but
they will be determined as a linear combination of the $a_i$s.
Indeed, we claim that the $a_i$s parametrize the most general R-symmetry \footnote{For a generalization of this sentence to singular horizons see subsection \ref{singhorizon}.}. The number of independent 
parameters in the trial R charge
is equal to the number of global $U(1)$ symmetries. We always have two 
global symmetries from the toric action and a number of baryonic symmetries 
equal to the number of three cycles. As shown in \cite{MSY}, the latter
is equal to $d-3$; 
each baryonic symmetry $B_a$ is indeed 
associated with a linear relation among the edges $V_i$
\begin{equation}
\sum_{i=1}^d B_i^a\, V_i=0
\label{linear}
\end{equation}   
and there are exactly $d-3$ such relations. In conclusion, we have
a total number of $d-1$ global $U(1)$ symmetries which matches the number
of independent parameters $a_i$.

Collecting all these pieces of information, we can propose the following
assignments of R-charges and multiplicities for the chiral fields
in the gauge theory:
\begin{itemize}
\item{Associate with each edge vector $V_i$ a chiral field with
trial R-charge $a_i$, with the constraint,
\begin{equation}
\sum_{i=1}^d a_i = 2 
\label{sommadue} 
\end{equation} }
\item{Call $C$ the set of all the unordered pairs of vectors in the $(p,q)$ web; we label an element of $C$ with the ordered indexes $(i,j)$, with the convention that  
the vector $v_i$ can be rotated to $v_j$ in the counter-clockwise direction
with an angle $\leq 180^o$. With our conventions $|\langle v_i, v_j \rangle|=\langle v_i, v_j \rangle$. Associate
with any element of $C$ the divisor 
\begin{equation}
D_{i+1}+D_{i+2}+ \ldots D_{j}
\end{equation}
and a type of chiral field in the field theory with multiplicity $\langle v_i, v_j \rangle$ and R-charge equal to $a_{i+1}+a_{i+2}+ \ldots a_{j}$. 
The indexes $i$, $j$ are always understood to be defined modulo $d$. For example in Figure \ref{pqweb} the field associated to the pair $(d,3)$ has R-charge $a_1+a_2+a_3$ and multiplicity $\langle v_d, v_3 \rangle$. 
The total number of fields is the sum of all the multiplicities:
\begin{equation}
E \equiv \sum_{(i,j) \in C}  |\langle v_i, v_j \rangle| 
\end{equation}
and thus reproduces formula~(\ref{numbers}).}
\end{itemize}

More generally,  we can assign global symmetry charges to all the fields. 
The algorithm is very similar to that for R-charges: 
\begin{itemize}
\item{
Assign global charges $a_i$ to the fields corresponding to vertices $V_i$.
The only difference is that now $a_i$ satisfy the relation:
\begin{equation}
\sum_{i=1}^d a_i =0
\label{sommazero}
\end{equation}  
}
\item{
The global charges of composite chiral fields are then: $a_{i+1}+a_{i+2} \ldots+a_j$ for the fields corresponding to $(i,j)$ in $C$.}
\end{itemize}

With a small abuse of notation, we will use the same letter $a_i$ for
R and global symmetries; in the first case they satisfy $\sum_{i=1}^d a_i=2$,
while in the latter $\sum_{i=1}^d a_i=0$.

Note that with the assignment  (\ref{sommazero}) 
we parametrize all the possible $d-1$ global symmetries, the $d-3$ baryonic
ones and the two flavor ones. We can explicitly identify the baryonic
symmetries as follows.
As shown in \cite{tomorrow}, the chiral fields 
associated with the edges $V_i$ have a charge under the baryonic symmetry $B_a$
equal to the coefficient $B^a_i$ in the linear relations~(\ref{linear}).
Notice that the baryonic charges of the fields associated with the edges
$V_i$ sum up to zero
\begin{equation}
\sum_{i=1}^d B^a_i=0
\label{sumbaryon}
\end{equation}
and therefore satisfy eq.~(\ref{sommazero})
as a consequence of the Calabi-Yau condition; the latter
requires that all the vectors $V_i$ lie on a plane which, in our conventions,
means that the first coordinate of all $V_i$ is $1$. In conclusion,
among the global symmetries, those
satisfying also the constraint (\ref{linear}) (with $a_i = B_i^a $) are 
the baryonic ones, the remaining two (for which there is not a natural basis, being mixed with baryonic symmetries \cite{tomorrow}) are the flavor ones.

We conjecture that for every Gorenstein toric singularity there exists a toric phase of the dual gauge theory where the R-charges and the multiplicities of all chiral fields can be computed with the algorithm above. This toric phase has generally the mi\-ni\-mal number of chiral fields (\ref{numbers}), as we have checked in many known cases. To be concrete look at Figure \ref{lpqr} corresponding to $L^{p,q;r,s}$. There are six kinds of fields: the four with charge $a_i$, fields with charge $a_3+a_4$ and others with charge $a_2+a_3$, but there are not for instance fields with charge $a_1+a_2$, since the region formed by $v_4$ and $v_2$ which includes $a_1$ and $a_2$ in the $(p,q)$ web has always an angle greater than $180^o$. Note that in general the number of different kinds of fields is $d(d-1)/2$, the number of elements of $C$. Note also that the R-charges of composite chiral fields can be written as sum of consecutive $a_i$s; since $P$ is convex the ordering of vectors $v_i$ in the $(p,q)$ web is always equal to the ordering of $v_i$ in $P$.  

%We shall also use the sets $C_h$, $h=1,2\ldots d$ which are subsets of $C$: a couple $(i,j)$ is in $C_h$ iff the R-charge of the corresponding chiral field is a sum $a_{i+1}+a_{i+2}+ \ldots a_{j}$ containing $a_h$. In practice $C_h$ is made up of all the couples $(i,j)$ such that the region $a_h$ in Figure \ref{pqweb} is contained in the angle $\leq 180^o$ generated by $v_i$ and $v_j$.

With this assignment, we have a trial central charge $a$ given by:
\begin{equation}
a=\frac{9}{32} \, \mathrm{tr}\, R^3 =\frac{9}{32} \left( F+ \sum_{(i,j) \in C}|\langle v_i, v_j \rangle| \,\, (a_{i+1}+a_{i+2}+\ldots a_{j}-1)^3 \right)
\label{aext}
\end{equation}   
Recall that $F$ is the double area of the polygon $P$ (\ref{numbers}).  The values of the R-charges $a_i$ can be found by (locally) maximizing this formula.
Note that this formula and the algorithm proposed above are obviously invariant under translations 
and $SL(2,\mathbb Z)$ transformations in the plane of $P$, since $\langle v_i, v_j \rangle$ are conserved, also
in sign \footnote{If the determinant is $-1$ all the signs are reversed and so relative orientations do not change.}.

We can make several checks of this proposal. First of all, it is  easy to
compare the proposal to the case where the quiver gauge theory is
explicitly known. Several examples are discussed in Section \ref{examples}.
In some cases, the fields and their multiplicity can be determined
by using mirror symmetry; this was done for the toric delPezzo in 
\cite{hananymirror} where the formula for the multiplicities based on the $(p,q)$ web first appeared. The multiplicity of the fields associated with
the edges $V_i$ was computed in \cite{tomorrow} and agrees with our proposal: 
\begin{equation}
\textrm{multiplicity of fields } (i-1,i) \in C= \langle v_{i-1}, v_i \rangle = \textrm{det} (V_{i-1},V_{i},V_{i+1})
\end{equation}
since the fields corresponding to the pair $(i-1,i)$ in $C$ have R-charge $a_i$; we have used that the first coordinates of $V_i$ are equal to 1. A proposal 
identical to ours was used in \cite{kru2} to determine the gauge theory
for the $L^{p,q,r}$ manifolds.

We can next study the consistency of our proposal with the general
properties of the $U(1)$ symmetries in our theories. First of all, we
must have
\begin{equation}
{\rm Tr} G=0 \, .
\end{equation}
where $G$ is a general R-charge or global symmetry charge. In particular
${\rm Tr} R={\rm Tr} B^a=0$. The proof of this formula is relatively easy and 
is reported in the Appendix. 
Another non trivial check
of our proposal is the proof, reported in the Appendix, that, for baryonic symmetries,
\begin{equation}
{\rm Tr} B_a^3=0
\end{equation}
This condition, which is true also for mixed baryonic symmetries, is a consequence
of the vanishing of the cubic t'Hooft anomaly for a baryonic symmetry.
This follows from the fact that on the stack of D3 branes in type IIB 
the baryonic symmetries are actually gauged. The counterpart of this statement
in the AdS dual is that cubic anomalies are computed from the Chern-Simons
terms in the five dimensional supergravity and no such term can contain three 
vector fields coming from reduction of the RR four-form \cite{wecht}.

The best check of the proposal is however the computation of the R-charges
at the fixed point using a-maximization and the comparison with
volumes of three cycles in $H$. 
Now that we have an algorithm to extract the field content of the gauge theory from the toric diagram, it is not difficult to write down an algorithm on a computer and check the agreement of a-maximization with Z-minimization on arbitrary large polytopes.  
The complete agreement of the a-maximization with the volume minimization of  \cite{MSY} will be discussed in details in the next Section, where a general 
analytic proof will be given.

We finish this Section by making some remarks about other toric phases of the same CFT with more chiral fields than the minimal phase presented above. 
In practical examples we often meet toric phases with the same trial central charge $a$ than the minimal phase;
these phases generally contain all the kinds of fields of the minimal phase, but with greater multiplicities.
In fact there are other possible
assignments of R-charges and multiplicities leading to the same $a$ charge.
%They presumably correspond to different toric phases of the same CFT.
For example, to each element in $C$ we may 
assign two different types of chiral fields, one associated with
the divisor 
\begin{equation}
C_{i,j}=D_{i+1}+D_{i+2}+...+D_{j}
\end{equation}
with R-charge $a_{i,j}=a_{i+1}+...a_{j}$, and a second one associated 
with the divisor \footnote{Recall that, in our conventions, the indexes
$i,j$ are always defined modulo $d$.}
\begin{equation}
\sum_{i=1}^d D_i-C_{i,j}=D_{j+1}+ D_{j+2}...+D_{i}
\end{equation}
with R-charge $a_{j+1}+...a_{i}=2-a_{i,j}$. If we assign multiplicities
$n_{i,j}$ and $\tilde n_{i,j}$ to the two types of fields with the constraint
\begin{equation}
n_{i,j} -\tilde n_{i,j}=|\langle v_i,v_j\rangle |
\end{equation}
it is easy to see that the equations ${\rm Tr} R={\rm Tr} B_a={\rm Tr} B_a^3=0$ are still satisfied. Moreover the expression for 
the trial central charge $a$ is unchanged. 
Indeed the contribution of the integers $n_{i,j}$ to the central charge 
cancels:
\begin{equation}
n_{i,j} (a_{i,j}-1)^3+ \tilde n_{i,j} (1- a_{i,j})^3\equiv |\langle v_i,v_j\rangle | (a_{i,j}-1)^3 \, .
\end{equation}
 The formula~(\ref{numbers}) for the number of chiral fields
is obviously no more satisfied. 
Each time a field is split and a new arbitrary
integer $n_{i,j}$ is introduced, the total numbers of fields increase.
Formula~(\ref{numbers}) is strictly valid for the minimal presentation.
%In this way, we have contructed infinite presentations of our CFT
%depending on arbitrary integers $n_{i,j}$, each one corresponding to a 
%different toric phase. 
We do not expect that for all arbitrary choices of $n_{i,j}$ and pairs $(i,j)$ there exists a non minimal toric phase with multiplicities of chiral fields described by this splitting mechanism, even though many known toric phases are characterized by multiplicities determined in this simple way.
In fact all the examples of (non minimal) toric phases considered in this paper are described by this splitting mechanism, and it would be interesting to know whether this is true in general. 
%Notice also that this splitting mechanism do not necessarily describe all the non minimal
%possible toric phases, see for example Model III of $dP_3$ in subsection \ref{dp3}. 
%Hopefully, our minimal presentation
%is the phase with the minimal number of fields.

\section{a-maximization is volume minimization}
\label{comparison}

For the purposes of the AdS/CFT correspondence, the R-charges of the
chiral fields have to be matched with the volumes of the three-cycles
bases $\Sigma_i$ of the corresponding divisors. 
In the previous Section we proposed a formula for computing the R-charges and the trial central charge $a$
directly from the toric diagram. Moreover in \cite{MSY} it was shown that all
the geometric information on the volumes can be extracted from the toric data,
through the process known as volume minimization (or Z-minimization),
without any explicit knowledge of the metric. The reason for that is the following: supersymmetric cycles are calibrated and the volumes can be extracted
only from the Kahler form on the cone. 
Therefore now it is possible to compare directly R-charges in the gauge theory
and volumes in the geometry, checking the correctness of the AdS/CFT predictions for every toric CY cone.
In this Section we discuss the equivalence of a-maximization and Z-minimization.

\begin{figure}
\centering
\includegraphics[scale=0.65]{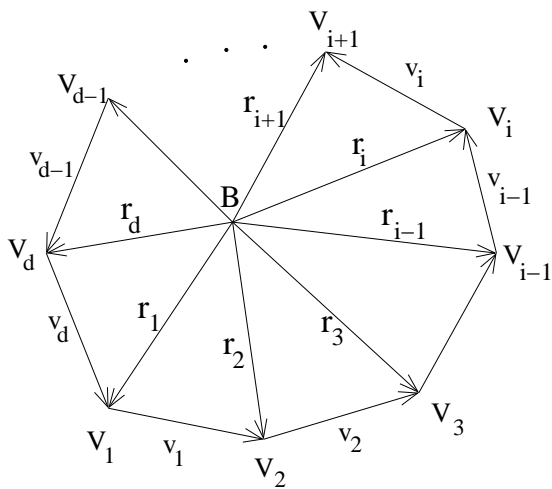}
\caption{The Reeb vector as a point $B$ inside the polygon $P$.}
\label{reeb}
\end{figure}

We start by reviewing the work of \cite{MSY} and reducing their formulas in the plane containing the convex polygon $P$.
The Reeb vector $K$ of a symplectic toric cone can be expanded in a basis $e_i$ for the $T^3$ 
effective action on the fiber:
 \begin{equation}\label{K}
K = \sum _{i=1} ^{3} b_i e_i
\end{equation} 
where the vector of coordinates $b=(b_1,b_2,b_3)$ lives inside the toric fan of the cone.
%In \cite{MSY} it was shown that there exists a unique choice of $b=\bar b$ for which it is possible to determine a CY metric over the symplectic toric cone and that the volume of the horizon $H$ and of the cycles $\Sigma_i$ are determined only by $\bar b$. 
The Reeb vector is associated
with an R-symmetry in the dual gauge theory; by varying the vector we
change the R-symmetry by mixing it with the 
global symmetries. From the geometrical point of view, the variation
of the Reeb vector changes the metric and the volumes.
For only one choice of vector $\bar b$ there exists a Calabi-Yau metric for the cone.
The vector $\bar b$ has $\bar b_1=3$ and can be determined through the minimization of a certain function Z of the variables $b_2$ and $b_3$ \cite{MSY}. We rephrase this process in the plane containing $P$ by writing $b=3(1,x,y)$ and allowing the point $B\equiv(x,y)$ to vary inside the convex polygon $P$: note in fact that $b$ is inside the fan. 
Define the functions:   
\begin{equation}\label{volumes}
{\rm Vol}_{\Sigma_i}(x,y)=\frac{2\pi^2}{9} \frac{\langle v_{i-1},v_i\rangle}{\langle r_{i-1},v_{i-1}\rangle \langle
 r_{i},v_{i}\rangle}\equiv \frac{2\pi^2}{9} l_i(x,y)
\label{li} 
\end{equation}
where $r_i$ is the plane vector going from $B$ to the vertex $V_{i}$ (see Figure \ref{reeb}). 
%These vectors evidently depend on $(x,y)$; note also that $r_{i+1}-r_i=v_i$.
As shown in \cite{MSY}, these are
the volumes of the base three-cycles associated with the divisors $D_i,i=1,...,d$.
Define also the function:
\begin{equation}\label{volumes2}
{\rm Vol}_H (x,y)=\frac{\pi}{6} \sum_{i=1}^d {\rm Vol}_{\Sigma_i}(x,y)
\end{equation}
which determines the total volume of the horizon $H$.
The two previous equations are just equations (3.25) and (3.26) of \cite{MSY}.
The function to minimize is just ${\rm Vol}_H (x,y)$ \footnote{This is the function $Z$ in \cite{MSY} up to a constant multiplicative factor.} and the position of the minimum $(\bar x, \bar y)$ gives the Reeb vector $\bar b=3(1,\bar x,\bar y)$ for the CY cone. It was proved in \cite{MSY} that such minimum exists and is unique. 

The values of ${\rm Vol}_H (\bar x,\bar y)$ and ${\rm Vol}_{\Sigma_i}(\bar x,\bar y)$ at the minimum are the total volume of $H$ and the volumes of $\Sigma_i$ to be compared with the central charge $a$ 
and the R-charges $a_i$ of the field theory through the AdS/CFT relations (\ref{central}) and (\ref{baryons}).
To facilitate this comparison we define the geometrical function:
\begin{equation}
a^{MSY}(x,y)=\frac{\pi^3}{4 {\rm Vol}_H(x,y)}
\label{aMSY}
\end{equation}
%(see equation (\ref{central})), 
and the functions:
\begin{equation}
f_i(x,y)= \frac{2 l_i(x,y)}{\sum_{j=1}^d l_j(x,y)}
\label{Rc}
\end{equation}
corresponding to the R-charges $R_i$ through equation (\ref{baryons}).
The process of Z-mi\-ni\-mi\-za\-tion can be restated as a maximization of $a^{MSY}(x,y)$ with $(x,y)$ varying in the interior of $P$.

On the other side of the correspondence we have the gauge theory with trial central charge $a$ which is a function of the $d$ variables $a_i$: 
\begin{equation}
 a(a_1,a_2,\ldots a_d)=\frac{9}{32} \left( F+ \sum_{(i,j) \in C}|\langle v_i, v_j \rangle| \,\, (a_{i+1}+a_{i+2}+\ldots a_{j}-1)^3 \right) 
\label{afield} 
\end{equation} 
We are considering a formal extension of the trial central charge to $\mathbb R^d$ defined by equation (\ref{aext}). This function has to be locally maximized with the constraint (\ref{sommadue}) (and $a_i>0$). To impose this constraint it is enough to introduce a Lagrange multiplier $\lambda$ and to extremize the function:
\begin{equation}
a(a_1,a_2,\ldots a_i)- \lambda (a_1+a_2\ldots +a_d-2)
\end{equation}
By deriving with respect to $a_i$ we get the conditions \footnote{This means that the gradient of the extended function $a$ in the local maximum is parallel to the vector $(1,\ldots 1)$. So to extremize $a$ it is enough to impose that the variations of $a$ along the $d-1$ vectors $S^a$ orthogonal to  $(1,\ldots 1)$ vanish:
\begin{equation}
\sum_{i=1}^d S^a_i \frac{\partial a}{\partial a_i}=0, \qquad \, \textrm{if} \quad \sum_{i=1}^d S^a_i=0
\label{global}
\end{equation}
But note that, in the language of Section \ref{geometry}, the space of $S^a$ is just the space of the $d-1$ global symmetries (compare with (\ref{sommazero})).}:
\begin{equation}
\frac{\partial a}{\partial a_i}=\lambda \qquad \quad i=1,\ldots d
\end{equation}

If we call $\bar a_i$ the values of $a_i$ at the local maximum, we have to prove that:
\begin{equation}
\begin{array}{l}
a^{MSY}(\bar x, \bar y) = a(\bar a_1, \bar a_2, \dots \bar a_d)\\[0.5em]
f_i(\bar x, \bar y)= \bar a_i \qquad i=1,\ldots d
\label{target}
\end{array}
\end{equation}
This is a highly non trivial check to perform:
a-maximization and Z-minimization use different functions and different trial charges;
it is not at all obvious why the result should be the same. 
First of all a-maximization is done on a total of $d-1$ independent
trial parameters while the volume minimization is done only on two parameters $(x,y)$. 
The trial central charge $a$ is a cubic polynomial in $a_i$, whereas 
$a^{MSY}$ is a rational function of $(x,y)$.
These parameters, in both cases, are somehow related to the possible global
symmetries: the Reeb vector in the geometry is connected to R-symmetries of the gauge theory 
and changing the position of $B$ in the directions $x$ and $y$ means adding to the R-symmetry
the two flavor global symmetries \footnote{Recall that flavor symmetries are mixed with baryonic ones,
so actually we are moving also in the space of baryonic symmetries.}. In any case, the volume minimization is done by moving only in a two dimensional
subspace of the set of global symmetries, while a-maximization is done
on the entire space. Fortunately, as often claimed in the literature,
a-maximization can be always performed on a two dimensional space of 
parameters related to flavor symmetries. Indeed, on general grounds, 
one can parametrize the trial R-symmetry as a contribution $R(X,Y)$ 
from the flavor symmetries plus a baryonic part
\begin{equation}
R=R(X,Y)+\sum_{a=1}^{d-3} h_a B_a
\label{rflavor}
\end{equation}
and the elimination of the variables $h_a$ is simple: imposing that the 
derivatives of $\mathrm{tr} R^3$ with
respect to $h_a$ vanish, one gets the equations 
\begin{equation}
{\rm Tr} R^2 B_a = 0 .
\label{ba}
\end{equation}
These conditions read
\begin{equation}
{\rm Tr} (R(X,Y)+\sum_{b=1}^{d-3} h_b B_b)^2 B_a = 0
\end{equation}
which is a linear system of $d-3$ equations in the $d-3$ variables $h_a$. Linearity in $h_a$
follows from the fact that the cubic mixed t'Hooft anomaly for baryonic symmetries is zero: ${\rm Tr}B_a^3=0$.
So one can solve for $h_a$ in function of $X$, $Y$ and substitute into the trial charge (\ref{rflavor}); 
the central charge $a$ is now a rational function only of $X$ and $Y$. So we have reduced a-maximization to a maximization over a set of two parameters.

In the previous argument, the choice of a basis of flavor and baryonic symmetries in 
(\ref{rflavor}) was quite arbitrary. In our specific case we can 
choose a more natural parameterization for 
the two dimensional space over which to reduce a-maximization. This space is just the space of coordinates $(x,y)$ of the plane where $P$ lies: consider the map from $\mathbb R^2$ to $\mathbb R^d$ given by
\begin{equation}
\begin{array}{l}
f:(x,y) \rightarrow (a_1,a_2,\ldots a_d)\\[1em]
\hspace{1.4em}(x,y) \rightarrow a_i= \displaystyle \frac{2 l_i(x,y)}{\sum_{j=1}^d l_j(x,y)}=f_i(x,y)
\label{map}
\end{array}
\end{equation} 
We claim that the local maximum $(\bar a_1,\ldots \bar a_d)$ of the a-maximization is found on the image $f(P)$ of the interior of $P$ under this map.
In fact it is not difficult to prove that the gradient of the trial central charge along the $d-3$ baryonic directions evaluated on $f(P)$ is always zero:
\begin{equation}
\sum_{i=1}^d B^a_i \frac{\partial a}{\partial a_i}_{|a_i=f_i(x,y)}=0
\label{derbaryon}
\end{equation} 
where $B^a$ is a baryon charge and where the equality holds for every $(x,y)$ in the interior of $P$. We give the general proof of (\ref{derbaryon}) in the Appendix. Note that equation (\ref{derbaryon}) is completely
equivalent to the condition (\ref{ba}) 
when the trial R-charge is evaluated with $a_i=f_i(x,y)$. Therefore we have clarified in which sense the baryonic symmetries decouple from the process of a-maximization.

At this point we have to compare the two functions $a^{MSY}(x,y)$ and the field theory trial central charge evaluated 
on the surface $f(P)$, which are two functions only of $(x,y)$. Remarkably one discovers that they are equal even before
maximization:
\begin{equation}
a(a_1,\ldots a_d)_{|a_i=f_i(x,y)}= a^{MSY}(x,y)
\label{equal}
\end{equation}
for every $(x,y)$ inside the interior of $P$. 
We give a general (still long) analytic proof of (\ref{equal}) in the Appendix.

%It is easy to check
%equation (\ref{equal}) with Mathematica. We have not yet found a simple
%analytical proof of equation (\ref{equal}). It is simple to check it in many concrete examples and we have some general evidences of its correctness. For example it is easy to prove that the two functions in (\ref{equal}) are both zero on the sides of $P$ and that their derivatives are equal on the sides. The explicit computation of ${\mbox Tr} R^3$ can be done with the methods described in the appendix, but the calculations are horribly long. We hope to find a 
%simple analytical proof in the future. 
 
%A general proof of (\ref{equal}) is still lacking; 
%yet we have checked it in many concrete examples and have some general evidences of its correctness. For example it is easy to prove that the two functions in (\ref{equal}) are both zero on the edges of $P$ and that their derivatives are equal on the edges. Equation (\ref{equal}), after explicitly substituting (\ref{Rc}) in (\ref{afield}), taking common denominator and simplifying factors of $\pi$ becomes:
%\begin{equation}
%F \left(\sum_{j=1}^d l_j \right)^3+\sum_{(i,j)\in C}|\langle v_i, v_j \rangle|\,
%(l_{i+1}+l_{i+2}\ldots l_j-l_{j+1}-l_{j+2}\ldots -l_i)^3=24 \left(\sum_{j=1}^d l_j \right)^2
%\label{geo}
%\end{equation}  
%which is a statement regarding only geometrical properties of polytopes $P$.

The proof of the equivalence of a-maximization and Z-minimization is now almost finished: we know that $a^{MSY}(x,y)$ has a unique maximum $(\bar x,\bar y)$ inside the polygon $P$. In this point we have for the field theory $a$:
\begin{equation}
\frac{\partial f_i}{\partial x_h}(\bar x,\bar y) 
\left(\frac{\partial a}{\partial a_i}\right)_{|a_i=f_i(\bar x, \bar y)} 
= \frac{d}{d x_h} a(f_1(x,y),\ldots f_d(x,y))_{|\bar x,\bar y}=
\frac{d}{d x_h} a^{MSY}(\bar x, \bar y)=0
\end{equation}
where $x_h$, $h=1,2$ is $x$ or $y$. So we see that, in the point $(\bar x, \bar y)$, also the two vectors:
\begin{equation}
\frac{\partial f_i}{\partial x}(\bar x,\bar y), \qquad \frac{\partial f_i}{\partial y}(\bar x,\bar y)
\end{equation}
belonging to the space of global symmetries (since $\sum_i f_i=2$) are orthogonal to the gradient of $a(a_1,\ldots,a_d)$.
Together with the $d-3$ baryon symmetries they span the whole $d-1$ space of global symmetries, thus proving (\ref{global}) in the point $(\bar x,\bar y)$. Therefore the extremum point for the trial central charge lies on the surface $f(P)$. One should also check that the Hessian matrix is negative definite to prove that this is a local maximum.
The agreement between the volumes of $\Sigma_i$ and the total volume with the R-charges $\bar a_i$ and the central charge in $(\bar x, \bar y)$ follows immediately from the parametrization (\ref{map}) and from (\ref{equal}).

\section{Examples}
\label{examples}
In this Section we provide various examples of our proposal using 
manifolds $H$ where it is possible to determine the dual gauge theory
explicitly. Needless to say, we find a remarkable agreement. 

\subsection{The $Y^{p,q}$ manifolds}
The superconformal theory dual to $AdS_5\times Y^{p,q}$ has been
determined in \cite{benvenuti}. 
The cone $C(Y^{p,q})$ determines
a polytope $P$ with vertices
\begin{equation}
(0,0)\, ,\qquad (1,0)\, \qquad  (0,p)\, \qquad (-1,p+q)
\label{ypqdiagram}
\end{equation}
with a $(p,q)$ web given by the vectors
\begin{equation}
(0,-1)\, ,\qquad (p,1)\, \qquad  (q,1)\, \qquad (-p-q,-1)
\end{equation}
With a toric diagram with four sides, we expect six different types of
fields corresponding to the number of pairs $(i,j)$. However, due to
the non-abelian isometry of the manifolds, there is an accidental 
degeneration. Our proposal and the comparison with the known
results is reported in Table 1 using the notations of 
\cite{benvenuti}. Recall that as usual $a_4=2-a_1-a_2-a_3$. With this assignment we would
perform the a-maximization on a three dimensional space of parameters.
The enhanced global symmetry allows to reduce the parameter space to
a two-dimensional one, as done in \cite{benvenuti}. Indeed,
in the a-maximization, $R$ can mix only with abelian symmetries 
\cite{intriligator}; we still have $d-3=1$ baryonic symmetries, but 
only one $U(1)$ flavor symmetry since the other is enhanced to $SU(2)$. 
In any event, without knowing about the $SU(2)$ symmetry we can perform
a-maximization on three parameters and discover at the end that $a_2=a_4$. 
In Table 1, four fields are associated with the four edges
of the fan. For $Y^{p,q}$ we obtain the fields $Y$,$Z$ and two copies
of the fields $U$ with the same multiplicity $p$: they combine to
give the $SU(2)$ doublet $U_\alpha$. The remaining two types of fields are
associated with the divisors $D_2+D_3$ and $D_3+D_4$, they have multiplicity
$q$ and combine to give the doublets $V_\alpha$.

\begin{center}
\begin{tabular}{|c|c|c|c|} \hline  
\raisebox{-0.28em}{$(i,j)\in C$} & \raisebox{-0.28em}{multiplicity} & \raisebox{-0.28em}{$U(1)_R$} & \raisebox{-0.28em}{fields} \\[0.5em] \hline 
\raisebox{-0.28em}{$(4,1)$} & \raisebox{-0.28em}{p+q} & \raisebox{-0.28em}{$a_1$} & \raisebox{-0.28em}{$Y$}  \\[0.5em] \hline
\raisebox{-0.28em}{$(1,2)$} & \raisebox{-0.28em}{p} & \raisebox{-0.28em}{$a_2$} & \raisebox{-0.28em}{$U$}  \\[0.5em] \hline
\raisebox{-0.28em}{$(2,3)$} & \raisebox{-0.28em}{p-q} & \raisebox{-0.28em}{$a_3$} & \raisebox{-0.28em}{$Z$}  \\[0.5em] \hline
\raisebox{-0.28em}{$(3,4)$} & \raisebox{-0.28em}{p} & \raisebox{-0.28em}{$a_4$} & \raisebox{-0.28em}{$U$}  \\[0.5em] \hline
\raisebox{-0.28em}{$(1,3)$} & \raisebox{-0.28em}{q} & \raisebox{-0.28em}{$a_2+a_3$} & \raisebox{-0.28em}{$V$}  \\[0.5em] \hline
\raisebox{-0.28em}{$(2,4)$} & \raisebox{-0.28em}{q} & \raisebox{-0.28em}{$a_3+a_4$} & \raisebox{-0.28em}{$V$}  \\[0.5em] \hline
\end{tabular} \\[1em]
Table 1: Charges and multiplicities for $Y^{p,q}$.
\end{center}

In the previous assignment $D_2+D_3$ has been chosen instead of $D_4+D_1$
because $\langle v_2,v_3\rangle >0$. A similar argument applies to
$D_3+D_4$. It is also easy to check that all the toric phases of $Y^{p,q}$ described
in \cite{phases} can be obtained in the way discussed at the end of
 Section \ref{geometry}
(cfr Table 1 in \cite{phases}). 

\subsection{The $L^{p,q;r,s}$ manifolds}

\begin{figure}
\centering
\includegraphics[scale=0.6]{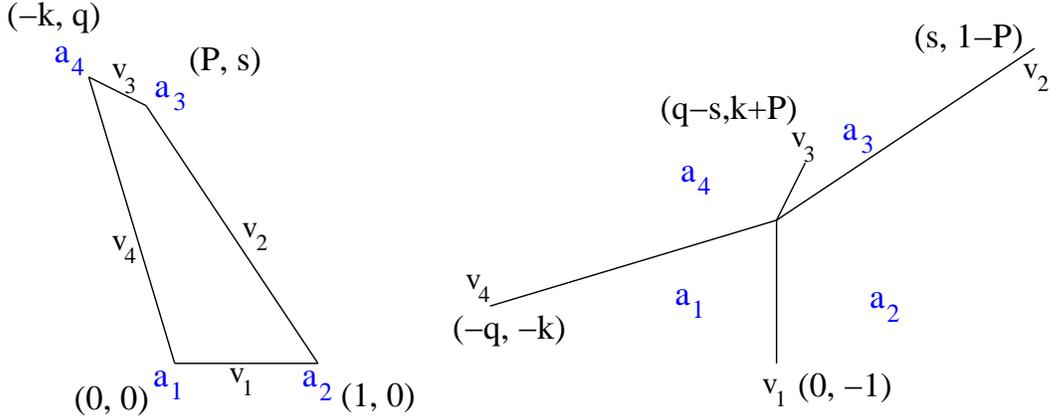}
\caption{The toric diagram and $(p,q)$ web for $L^{p,q;r,s}$}
\label{lpqr}
\end{figure}

The superconformal theory dual to $AdS_5\times L^{p,q;r,s}$ has been
determined in \cite{kru2,noi,tomorrow}. The cone $C(L^{p,q;r,s})$ determines
a polytope $P$ with vertices
\begin{equation}
(0,0)\, ,\qquad (1,0)\, \qquad  (P,s)\, \qquad (-k,q)
\end{equation}
where $k$ and $P$ are determined through the Diophantine equation
\begin{equation}
r-k s-P q=0
\end{equation}
Recall that $p+q=r+s$. As explained in \cite{noi}, we can always choose
$p\le r\le s\le q$ without any loss of generality.
The $(p,q)$ web is given by the vectors
\begin{equation}
(0,-1)\, ,\qquad (s,1-P)\, \qquad  (q-s,k+P)\, \qquad (-q,-k)
\end{equation}
The toric diagram and $(p,q)$ web for $L^{p,q;r,s}$ are reported in Figure \ref{lpqr}.

\vspace{0.7em}

\begin{center}
\begin{tabular}{|c|c|c|c|} \hline  
\raisebox{-0.28em}{$(i,j)\in C$} & \raisebox{-0.28em}{multiplicity} & \raisebox{-0.28em}{$U(1)_R$} & \raisebox{-0.28em}{fields} \\[0.5em] \hline 
\raisebox{-0.28em}{$(4,1)$} & \raisebox{-0.28em}{q} & \raisebox{-0.28em}{$a_1$} & \raisebox{-0.28em}{$Y$}  \\[0.5em] \hline
\raisebox{-0.28em}{$(1,2)$} & \raisebox{-0.28em}{s} & \raisebox{-0.28em}{$a_2$} & \raisebox{-0.28em}{$\tilde W$}  \\[0.5em] \hline
\raisebox{-0.28em}{$(2,3)$} & \raisebox{-0.28em}{p} & \raisebox{-0.28em}{$a_3$} & \raisebox{-0.28em}{$Z$}  \\[0.5em] \hline
\raisebox{-0.28em}{$(3,4)$} & \raisebox{-0.28em}{r} & \raisebox{-0.28em}{$a_4$} & \raisebox{-0.28em}{$X$}  \\[0.5em] \hline
\raisebox{-0.28em}{$(1,3)$} & \raisebox{-0.28em}{q-s} & \raisebox{-0.28em}{$a_2+a_3$} & \raisebox{-0.28em}{$W$}  \\[0.5em] \hline
\raisebox{-0.28em}{$(2,4)$} & \raisebox{-0.28em}{q-r} & \raisebox{-0.28em}{$a_3+a_4$} & \raisebox{-0.28em}{$\tilde X$}  \\[0.5em] \hline
\end{tabular} \\[1em]
Table 2: Charges and multiplicities for $L^{p,q;r,s}$.
\end{center}

The toric diagram has four sides, and we expect six different types of
fields corresponding to the number of pairs $(i,j)$. In this case 
the isometry is $U(1)^3$ and we don't expect any degeneration.
Our proposal and the comparison with the known
results is reported in Table 2 using the notations of 
\cite{noi}.
Recall that as usual $a_4=2-a_1-a_2-a_3$. The same analysis was performed in \cite{kru2}.

\subsection{The $X^{p,q}$ manifolds}
It is interesting to check the case of the manifolds $X^{p,q}$ discussed
in \cite{hananyX}. These correspond to toric cones with five facets
which can be blown down to the cones over $Y^{p,q}$. The corresponding
gauge theory can be determined by an inverse Higgs mechanism \cite{hananyX}.
The general assignment of R-charges and the a-maximization has not been 
performed in the literature except for particular $p$ and $q$;
therefore this model is an interesting laboratory.

\begin{figure}
\begin{minipage}[t]{0.4\linewidth}
\vspace{0pt}
\centering
\includegraphics[scale=0.56]{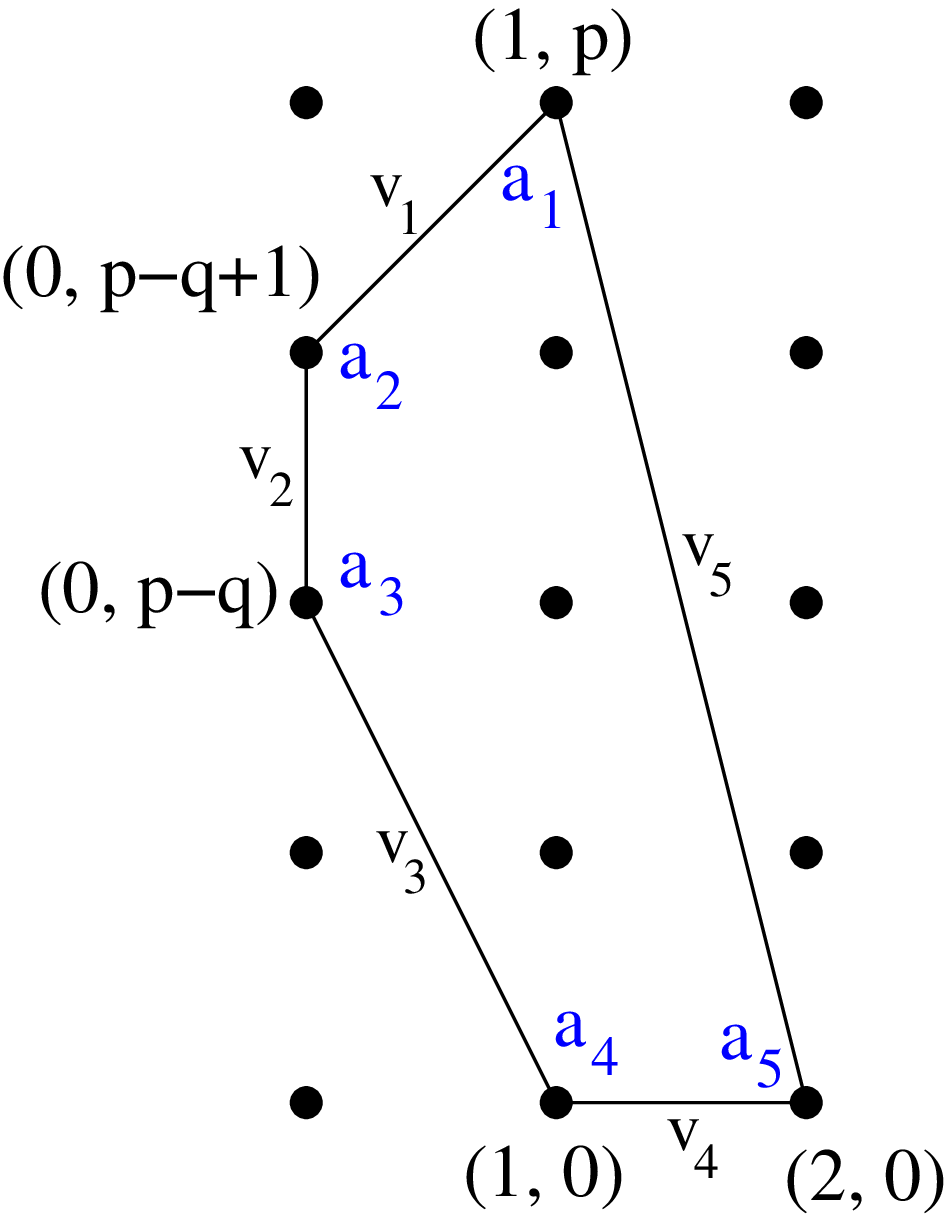}
\caption{Toric diagram for $X^{p,q}$. We draw the case $X^{4,2}$.}
\label{toricxpq}
\end{minipage}%
~~~~~~\begin{minipage}[t]{0.58\linewidth}
\vspace{65pt}
\centering
\includegraphics[scale=0.56]{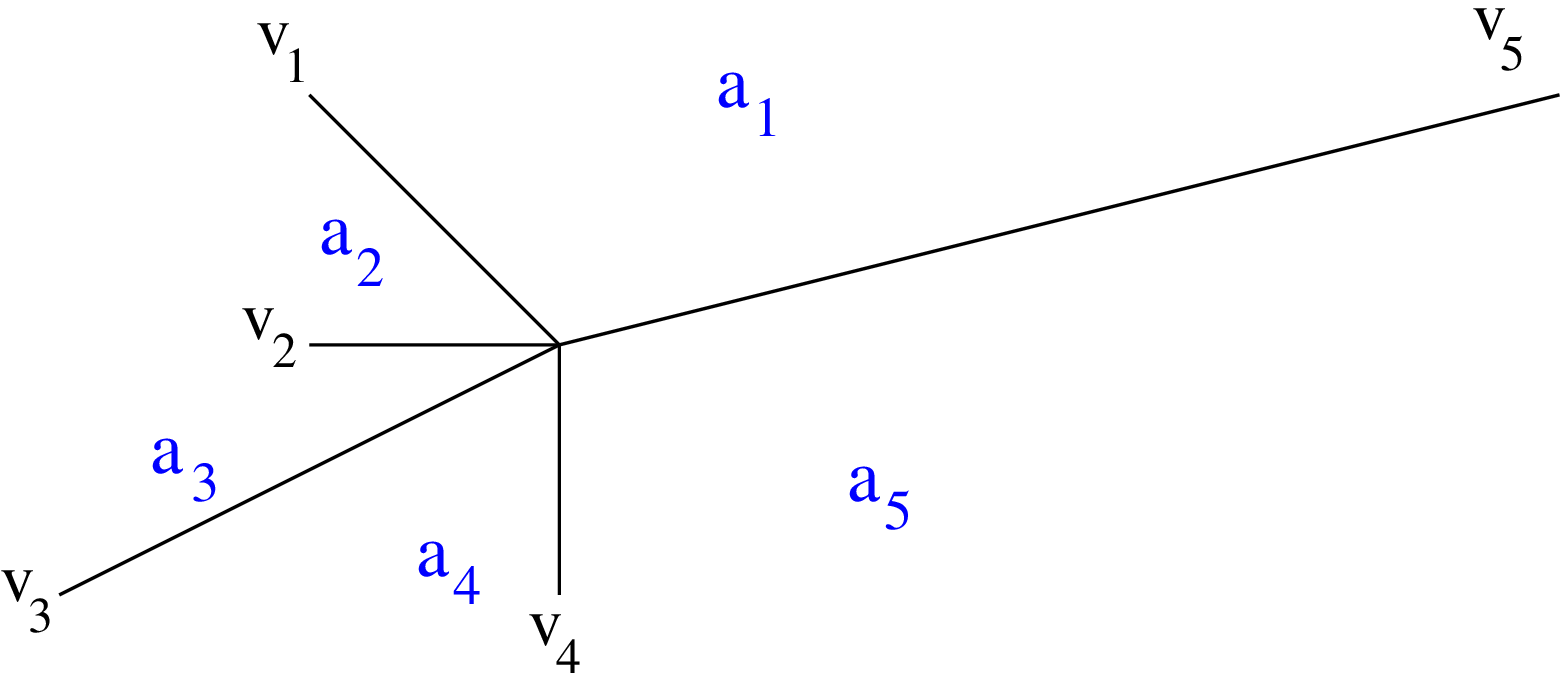}
\vspace{14pt}
\caption{The $(p,q)$ web for $X^{p,q}$.}
\label{pqwebxpq}
\end{minipage}%
\end{figure}

The toric diagram is given by (see Figure \ref{toricxpq}): 
\begin{equation}
(1,p)\, ,\qquad (0,p-q+1)\, \qquad  (0,p-q)\, \qquad (1,0)\, \qquad (2,0)
\label{diagramxpq}
\end{equation}
and the $(p,q)$ web is given by the vectors $v_i$ (see Figure \ref{pqwebxpq}):
\begin{equation}
(-q+1,1)\, ,\qquad (-1,0)\, \qquad  (-p+q,-1)\, \qquad (0,-1)\, \qquad (p,1)
\end{equation}
With a toric diagram with five sides, we expect ten different types of
fields corresponding to the number of pairs $(i,j)$. 
Our proposal is reported in Table 3.
Recall that as usual $\sum_i a_i=2$ for an R-symmetry. We have four independent 
parameters because there are now two baryonic symmetries.

\begin{center}
\begin{tabular}{|c|c|c|} \hline  
\raisebox{-0.28em}{$(i,j)\in C$} & \raisebox{-0.28em}{multiplicity} & \raisebox{-0.28em}{$U(1)_R$} \\[0.5em] \hline 
\raisebox{-0.28em}{$(5,1)$} & \raisebox{-0.28em}{p+q-1} & \raisebox{-0.28em}{$a_1$} \\[0.5em] \hline
\raisebox{-0.28em}{$(1,2)$} & \raisebox{-0.28em}{1} & \raisebox{-0.28em}{$a_2$} \\[0.5em] \hline
\raisebox{-0.28em}{$(2,3)$} & \raisebox{-0.28em}{1} & \raisebox{-0.28em}{$a_3$} \\[0.5em] \hline
\raisebox{-0.28em}{$(3,4)$} & \raisebox{-0.28em}{p-q} & \raisebox{-0.28em}{$a_4$} \\[0.5em] \hline
\raisebox{-0.28em}{$(4,5)$} & \raisebox{-0.28em}{p} & \raisebox{-0.28em}{$a_5$} \\[0.5em] \hline
\raisebox{-0.28em}{$(1,3)$} & \raisebox{-0.28em}{p-1} & \raisebox{-0.28em}{$a_2+a_3$} \\[0.5em] \hline
\raisebox{-0.28em}{$(2,4)$} & \raisebox{-0.28em}{1} & \raisebox{-0.28em}{$a_3+a_4$} \\[0.5em] \hline
\raisebox{-0.28em}{$(1,4)$} & \raisebox{-0.28em}{q-1} & \raisebox{-0.28em}{$a_2+a_3+a_4$} \\[0.5em] \hline
\raisebox{-0.28em}{$(5,2)$} & \raisebox{-0.28em}{1} & \raisebox{-0.28em}{$a_1+a_2$} \\[0.5em] \hline
\raisebox{-0.28em}{$(3,5)$} & \raisebox{-0.28em}{q} & \raisebox{-0.28em}{$a_4+a_5$} \\[0.5em] \hline
\end{tabular} \\[1em]
Table 3: Charges and multiplicities for $X^{p,q}.$
\end{center}

\begin{figure}[ht]
\begin{minipage}[t]{\linewidth}
\begin{minipage}[t]{0.3\linewidth}
\vspace{0pt} 
\includegraphics[scale=0.67]{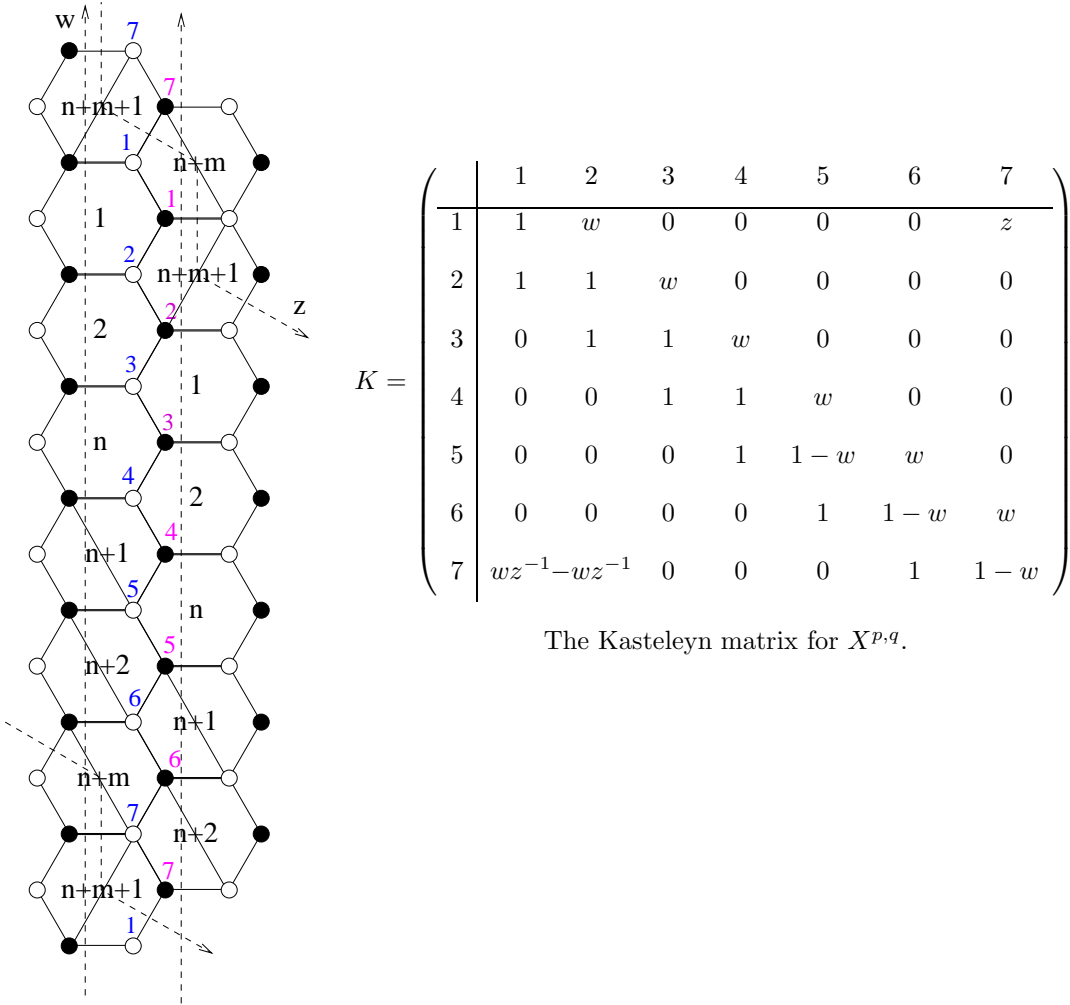}
\vspace{0pt} \hspace{-10em}
\caption{The brane tiling for $X^{p,q}$.}
\label{xpq}
\end{minipage} 
\begin{minipage}[t]{0.5\linewidth}
\vspace{0pt} 
\vspace{5em}
\centering
\[
K= \left(
\begin{array}{c|c@{\hspace{-0.5pt}}c@{\hspace{1em}}c@{\hspace{2em}}c@{\hspace{1.5em}}ccc}
    & 1 & 2 & 3 & 4 &  5 & 6 & 7 \\[0.5em] \hline 
 1  & 1 & w & 0 & 0 &  0  &  0  &  z \\[1em]
 2  & 1 & 1 & w & 0 &  0  &  0  &  0 \\[1em]
 3  & 0 & 1 & 1 & w &  0  &  0  &  0 \\[1em]
 4 & 0 & 0 & 1 & 1 &  w  &  0  &  0 \\[1em]
 5  & 0 & 0 & 0 & 1 & 1-w &  w  &  0 \\[1em]
 6  & 0 & 0 & 0 & 0 &  1  & 1-w &  w \\[1em]
 7  &wz^{-1}& -wz^{-1} & 0 & 0 &  0  & 1 & 1-w  \\[0.5em]
\end{array}
\right)
\]
\vspace{1em}
\verb| | \hspace{5em} The Kasteleyn matrix for $X^{p,q}$. 
\end{minipage}
\end{minipage}
\end{figure}

We can explicitly determine the gauge theory and assign the R-charges
to bi-fundamental fields. This can be done more efficiently using the
brane tiling description of the $X^{p,q}$ theory. We refer to \cite{dimers}
for a detailed discussion of the brane tiling. Here we use the method
we employed for the $L^{p,q;r,s}$ manifolds in \cite{noi}. The tiling
for $X^{p,q}$ is pictured in Figure \ref{xpq}. 
Similarly to $Y^{p,q}$ theories, the dimer configuration of $X^{p,q}$
can be obtained using only one column of $n$ hexagons, and $m+1$ 
consecutive cut hexagons. The horizontal identification has a shift $k=1$,
as for $Y^{p,q}$. The main difference is that for $X^{p,q}$ the last cut
hexagon has a cut in the opposite direction than the other $m$ cuts.

To fit the number of fields, gauge groups and superpotential terms for $X^{p,q}$
we must choose: $n=2q-1$, $m=p-q$.
We report also the general form of the Kasteleyn matrix,
with vertices numbered in the same way as in \cite{noi} (see Figure \ref{xpq}).
The determinant of $K$ is then:
\begin{equation}
\mathrm{det}\, K = 1+ z + z^{-1}w^{n+m+1}+z^{-1}w^{n+m}+ w^{\frac{n+1}{2}+m}+\ldots
\end{equation}
where we have not been careful about signs and the omitted terms are powers
of $w$ with lower exponent. In the plane $(z,w)$ one gets the toric diagram:
\begin{equation}
(0,0)\, ,\qquad (1,0)\, \qquad  (0,p)\, \qquad (-1,p+q)\, \qquad (-1,p+q-1)
\label{xpqdue}
\end{equation}
Translating by $(0,-p)$, applying the $SL(2,\mathbb Z)$ transformation $((1,0),(-p,-1))$ and translating by $(1,0)$, one recovers the equivalent diagram
(\ref{diagramxpq}). This shows that the dimer configuration reproduces the geometry. By comparing~(\ref{xpqdue}) with (\ref{ypqdiagram}), it is manifest that
the cone $C(X^{p,q})$ can be obtained by blowing up  $C(Y^{p,q})$.
Using the tiling in Figure \ref{xpq} and the algorithm
described in \cite{noi} \footnote{the four independent symmetries are determined by the assignments $v^1$, $v^2$, $v^3$ as in Appendix A.2 of \cite{noi} plus an assignment built as a second ``cycle'' starting from the cut hexagon at position $m+1$. An alternative and more general method for determining the distribution of R-charges on the dimer is described in subsection \ref{distribution}.} we can find the different types of fields and their distribution on the tiling
in the general case. The agreement with our proposal given in Table 3 is complete.

\subsection{Assigning R-charges on the dimer: a general conjecture}
\label{distribution}

In this subsection we propose, and check also on specific examples, a general conjecture to assign R-charges to chiral fields. 

In \cite{dimers} it was suggested a natural one to one correspondence between auxiliary fields in the Witten sigma model associated with a quiver theory and perfect matchings of the dimer configuration. Recall that a perfect matching of a bipartite graph is a choice of links such that every white and black vertex is taken exactly once. 
We will concentrate on theories for which the multiplicities of the auxiliary fields in the associated Witten sigma model \footnote{not to be confused with the multiplicity of the ``real'' fields in the gauge theory associated to the vertex of the toric diagram.} corresponding to vertices of the toric diagram are all equal to one. That is we consider dimer configurations with only one perfect matching corresponding to each vertex of the toric diagram.
%(that can be computed from the dimer through the determinant of the Kasteleyn matrix \cite{dimers}).
This is always true in all known theories we considered and we think this may be true also in general.
 %(at least for the toric phase with minimal number of fields we are dealing in this paper).

In fact not only there exist many equivalent descriptions (dimer configurations) of the same physical theory, generally connected by Seiberg dualities, but there are also dimer configurations that do not have any AdS/CFT dual. As an example consider the dimers that can be built using only one column of $n$ hexagons and $m$ (consecutive) cut hexagons as in \cite{noi}. In that paper it was pointed out that, using an horizontal identification with shift $k=1$, one can obtain the whole family of $Y^{p,q}$ theories with the choice $n=2q$ and $m=p-q$. 
Note that $n$ is always even, and the toric diagram of $Y^{p,q}$ is:
\begin{equation}
(0,0), \quad (1,0), \quad (0,n/2+m), \quad (-1,n+m) \qquad \textrm{for $n$ even}
\end{equation}
For these configurations the number of perfect matchings associated to any vertex $V_i$ is always one, as it is easy to prove from the general expression of the Kasteleyn matrix reported in \cite{noi}. Moreover these configurations
survive the test of the equivalence between a-maximization and Z-minimization.

Instead if we build tilings with an odd number $N$ of normal hexagons and $M$ cut hexagons (shift again k=1) we get surprising results.
The toric diagram is now given by:
\begin{equation}
(0,0), \quad (1,0), \quad (0,\frac{N-1}{2}+M), \quad (-1,N+M) \qquad \textrm{for $N$ odd}
\end{equation}
as one can see from the Kasteleyn matrix. Note that, up to auxiliary fields multiplicities, we get the same toric configuration if we choose:
\begin{equation}
N=n-1 \qquad M=m+1
\end{equation}   
but with N odd there is a vertex of the toric diagram (precisely $(0,(N-1)/2+M)$) having more than one corresponding perfect matching, as one can see again from the Kasteleyn matrix. Moreover it is easy to see that the theories corresponding to configurations $(N,M,k=1)$ with $N$ odd do not match the Z-minimization results of the corresponding toric diagrams. These quiver theories do not have a conformal fixed point satisfying the unitary bounds. In fact it is easy to convince oneself that they have only two \footnote{With $N$ odd the cycle described in Appendix A.2 of \cite{noi} to build the third charge do not cover all the cut hexagons.} global symmetries instead of three $U(1)$ symmetries of $Y^{p,q}$  (one of these $U(1)$ is however enlarged to $SU(2)$ for $Y^{p,q}$). 
The trial R-charges associated with some fields (those corresponding to the cuts of the hexagons) are zero and this violates the unitary bound since the
corresponding gauge invariant dibaryon operators would have dimension zero. 
In this way, we have built an infinite family of quiver gauge theories, that can be represented with dimer configurations, but cannot have any geometric AdS/CFT dual. We analyzed some other cases of theories without a geometric dual by varying also $k$, and always found that such theories have at least a vertex of the toric diagram with number of perfect matchings associated greater than one.
We conjecture that the request of having only a perfect matching corresponding to each vertex of the toric diagram is necessary for the existence of an AdS/CFT dual, but this statement should be further studied.
In the following we only consider theories that satisfy such request.

Our conjecture is that it is possible to assign R charges (or global charges) once the perfect matchings corresponding to the vertices of the toric diagram are known \footnote{We consider here smooth horizons for which the edges of the toric diagram do not pass through integer points, for an extension of this conjecture also to non smooth horizons see subsection \ref{singhorizon}.}. The method is simple: assign R-charge (or global charge) $a_i$ to the perfect matching corresponding to the vertex $V_i$ of the toric diagram. The charges $a_i$ satisfy (\ref{sommadue}) if they are R-charges or (\ref{sommazero}) if they are global charges. The (R-)charge of a link in the dimer configuration is then the sum of all (R-)charges of the perfect matchings (corresponding to vertices of the toric diagram) to which the link belongs. 

We have checked in many known cases that this method works, also in different toric phases of the same theory. For phases with the minimal number of fields it reproduces our formula for the multiplicities of the different kinds of fields.
For example it is not difficult to extract the perfect matchings associated to vertices of the $X^{p,q}$ theories from the Kasteleyn matrix reported in the previous subsection. And then one can check that the distribution of R-charges in the dimer obtained with the method proposed is a good distribution, that is one verifies that at every vertex the sum of R-charges is 2 (invariance of the superpotential) and for every face the sum of R-charges is equal to the number of edges minus 2 (beta functions equal to zero). 
We give other explicit examples of this method in the following subsection.

It would be interesting to check whether this method works in general. 
Obviously the invariance of the superpotential is guaranteed, since every perfect matching takes every vertex once and the sum of the $a_i$ is 2. It would be necessary also to prove the condition for faces (zero beta functions).
%, which is probably connected to the fact that we are considering perfect matchings associated to the vertices.

In the toric phases with minimal number of fields the method for computing multiplicities of fields described in Section \ref{geometry} should hold. Every perfect matching is made up with
 $V/2$ links, where $V$, the number of vertices in the dimer configuration, is computed in minimal phases from the toric diagram as $V=E-F$. The method proposed in this subsection implies that there are exactly $V/2$ fields containing 
%(maybe summed with other charges) 
the charge $a_1$, and the same is true for every $a_i$, $i=1,\ldots d$. Consistence with our formulas for computing multiplicities from the $(p,q)$ web requires that the sum of multiplicities of all fields containing $a_i$ is equal to $V/2$ independently from $i$. This is true and is proved in Appendix \ref{useful}.  

As a final remark, let us remind that in \cite{dimers} it was discovered that a chiral field (a link in the dimer) in the gauge theory can be written as the product of all auxiliary fields associated to perfect matchings to which the field belongs, and not only to perfect matchings corresponding to vertices of the toric diagram. Hence we have claimed that only the perfect matchings associated to vertices are charged under R or global symmetries, whereas other perfect matchings have charges equal to zero.

\subsection{The toric del Pezzo 3}
\label{dp3}

\begin{figure}
\begin{minipage}[t]{0.48\linewidth}
\vspace{0pt}
\centering
\includegraphics[scale=0.68]{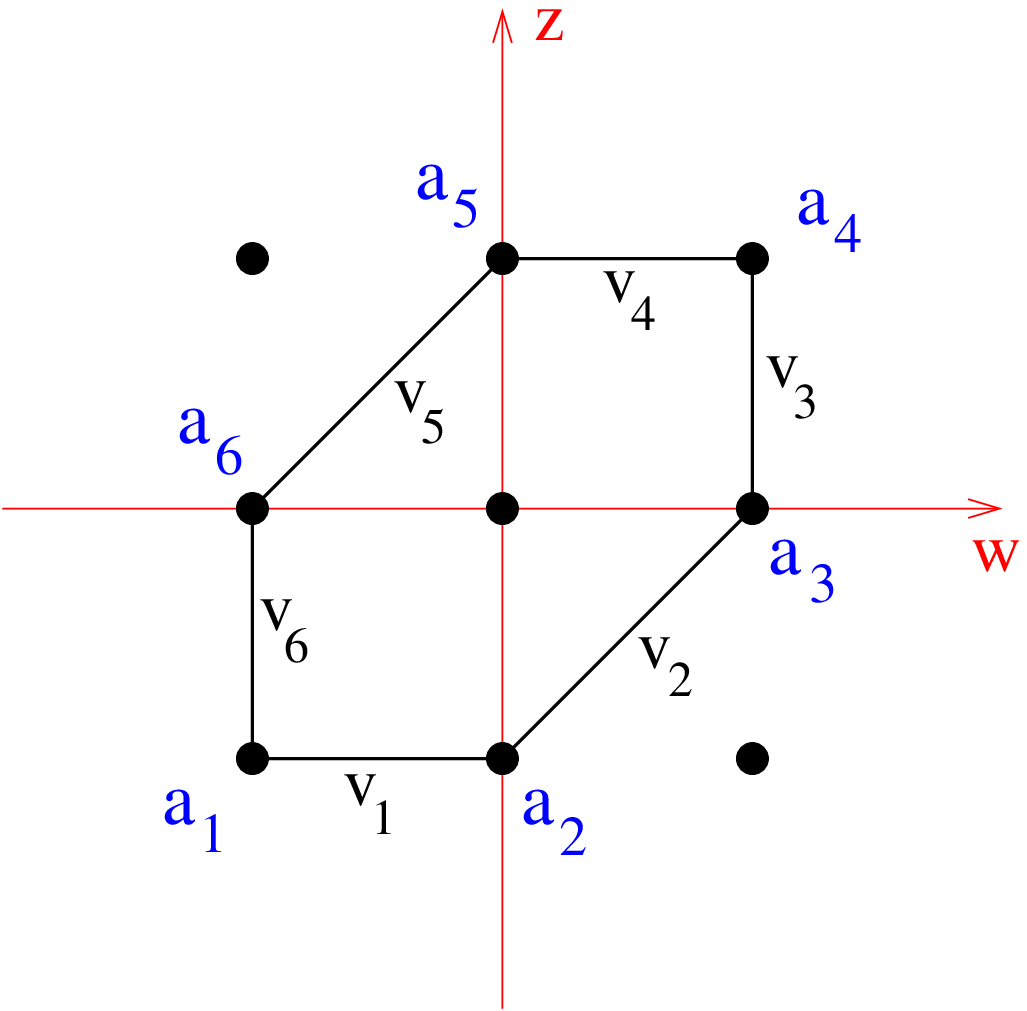}
\caption{Toric diagram for $dP_3$.}
\label{toricdp3}
\end{minipage}%
~~~~~~\begin{minipage}[t]{0.48\linewidth}
\vspace{23pt}
\centering
\includegraphics[scale=0.9]{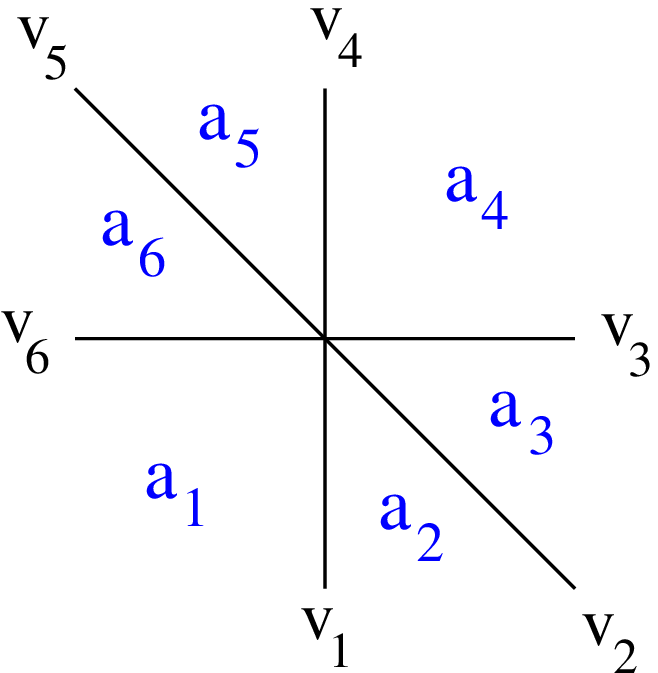}
\caption{The $(p,q)$ web for $dP_3$.}
\label{pqwebdp3}
\end{minipage}%
\end{figure}

In this subsection we consider the example of the theories associated with the
 complex cone over $dP_3$. This toric manifold is interesting since its toric diagram 
has six edges and four different toric phases are known. All the corresponding quivers are given in \cite{dimers}. 
We draw in Figures \ref{toricdp3} and \ref{pqwebdp3} the toric diagram and $(p,q)$ web for $dP_3$; we also show the assignment of charges $a_i$ in our conventions. Remember that for R-charges the sum
 of all $a_i$ is equal to 2.  

\begin{figure}
\centering
\includegraphics[scale=0.5]{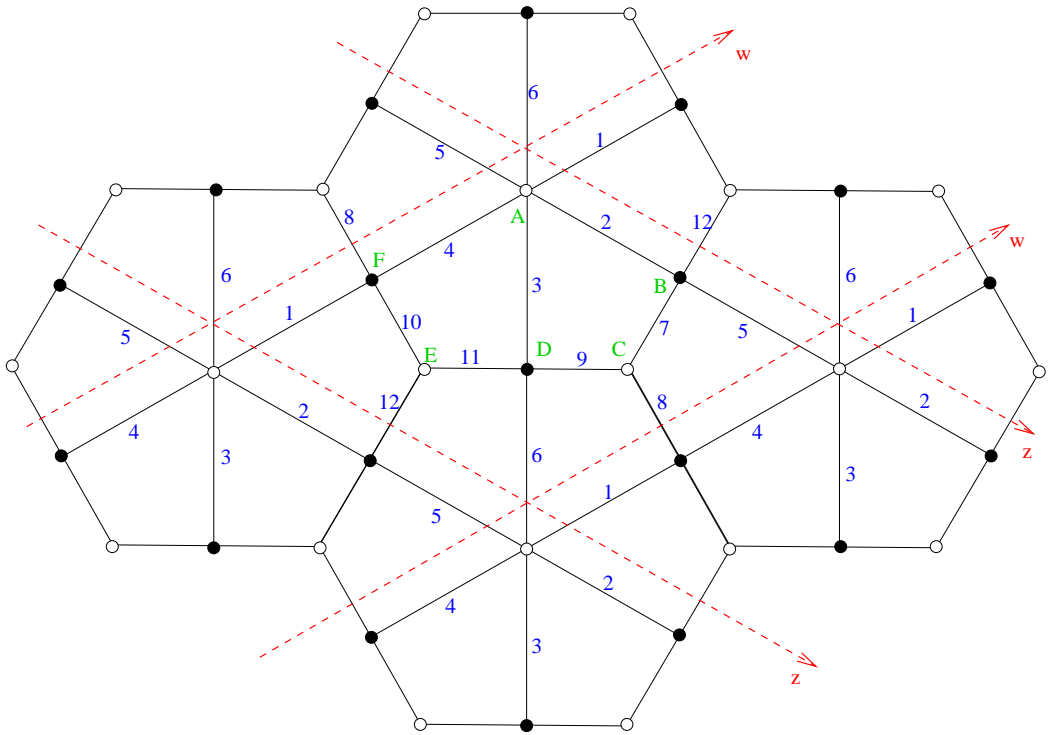}
\caption{The dimer configuration for $dP_3$, Model I.}
\label{dp3m1}
\end{figure}

The area of the toric diagram is 3, and therefore the number of gauge groups is $F=6$.
Model I of $dP_3$ has 12 fields $E=12$ and hence $V=6$ terms in the superpotential. This model has the least number of fields among the toric phases of $dP_3$, in agreement with equations (\ref{numbers}). We draw the dimer configuration for Model I in Figure \ref{dp3m1}; we label the chiral fields $X_i$ with numbers $i=1,\ldots 12$ typed in blue and vertices with letters $A,B,\ldots F$. The
identification of faces is as in \cite{dimers}.

To compute the R-charges of the theory we can use the method suggested in the previous subsection; first we have to know the perfect matchings associated to the vertices. A fast way to compute them is by writing the determinant of a modified Kasteleyn matrix:
\begin{equation}
K= 
\begin{array}{c|ccc}
 & B & D & F \\ \hline
A & X_2+ w X_5 & X_3 - w z X_6 & X_4+z X_1 \\
C & X_7 & -X_9 & -w^{-1}X_8 \\
E & -z^{-1}X_{12} & -X_{11} & X_{10} 
\end{array}
\end{equation}
where we have written for every field not only the usual weight in function of $w$ and $z$ \cite{dimers}, but also the name of the field itself. Note that it is not necessary to be careful about signs.
The coefficient of $w^i z^j$ in the expression of $\mathrm{det} K$ gives the perfect matching(s) associated with the point at position $(i,j)$ in the plane of the toric diagram. 
%This is obviously always true since the determinant is a sum of products of the entries of $K$ where each row and column index (white and black vertex) is taken exactly once in each product. 
So we find that the perfect matchings associated with the vertices are:
\begin{eqnarray}
a_1 & \rightarrow & X_3 X_8 X_{12} \nonumber \\ 
a_2 & \rightarrow & X_4 X_9 X_{12} \nonumber \\
a_3 & \rightarrow & X_5 X_9 X_{10} \nonumber \\
a_4 & \rightarrow & X_6 X_7 X_{10} \nonumber \\
a_5 & \rightarrow & X_1 X_7 X_{11} \nonumber \\ 
a_6 & \rightarrow & X_2 X_8 X_{11} \nonumber \\    
\end{eqnarray}
where on the left we have written the R-charge associated with 
the vertex/perfect matching.
We can then compute the R-charges of the fields $X_i$ as described in the previous subsection by summing all the charges of the vertex perfect matchings to which a field
 $X_i$ belongs. We thus get the following table for R-charges:                                              
\begin{equation}
\begin{array}{cccccccccccc}
X_1 & X_2 & X_3 & X_4 & X_5 & X_6 & X_7 & X_8 & X_9 & X_{10} & X_{11} & X_{12}\\
a_5 & a_6 & a_1 & a_2 & a_3 & a_4 & a_4+a_5 & a_1+a_6 & a_2+a_3 & a_3+a_4 & a_5+a_6 & a_1+a_2 \nonumber
\end{array}
\label{modI}
\end{equation}
We have found five independent trial R-charges (there is relation (\ref{sommadue}) among the $a_i$), and indeed it is not difficult to show that they are the correct ones, for example by writing the matrix $C_{ij}$ as in Appendix A.2 of \cite{noi}. 

Note that the multiplicities (equal to 1 for $dP_3$) and the kinds of different fields just found are in agreement with the general formula we propose in this paper. So we recognize in Model I the minimal toric phase of $dP_3$ for which the formulae proposed in this paper strictly hold.

There are three other phases of $dP_3$ with more than 12 fields. We have performed a similar analysis also for these phases, taking the dimer diagrams from \cite{dimers}.
We do not report here all the calculations, but make some useful comments. First of all we have checked that one can use the algorithm described in the previous subsection to determine the R-charges; this is an efficient and fast algorithm. 

Model II, III and IV fit in the general analysis at the end of Section \ref{geometry}.  Model II and III of $dP_3$ have $F=6$, $E=14$, $V=8$. They both have the same distribution of fields: there are all the fields that appeared in Model I with the same R-charges (\ref{modI}) plus two other fields: one has R-charge $a_3+a_4+a_5$ and the other $a_1+a_2+a_6$. Their contribution to the trial central charge $a$ cancels:
\begin{equation}
\left( a_3+a_4+a_5-1 \right)^3 +\left( a_1+a_2+a_6-1 \right)^3=0
\end{equation}
because of (\ref{sommadue}). So the trial a charge to maximize is the same as in Model I.\footnote{Note that there may exist other parametrizations of trial R-charges. For example in Model III, one can find an equivalent distribution interchanging $a_1$ and $a_4$: this still satisfies the linear constraints from the vanishing of beta functions and conservation of superpotential. The expression of the trial central charge to maximize is different, but the results at the maximum, where $a_1=a_4$, are the same.  This is due to the high degree of symmetry of the toric diagram of $dP_3$.}
Model IV of $dP_3$ has $F=6$, $E=18$, $V=12$. There are all the fields appearing in Model I plus the six fields with R-charge: $a_3+a_4+a_5$, $a_1+a_2+a_6$, 
 $a_5+a_6+a_1$,  $a_2+a_3+a_4$, $a_1+a_2+a_3$, $a_4+a_5+a_6$. Again their contribution to the trial central charge cancels.

%Model III is more interesting; it has $F=6$, $E=14$, $V=8$. The R charges of the 14 fields are:
%\begin{equation}
%\begin{array}{l}
%\begin{array}{ccccccccccc}
%a_1, & a_2, & a_3, & a_4, & a_5, & a_6, & a_1+a_3, & a_1+a_5, & a_2 +a_4, & a_6+a_4, & a_2+a_3,  
%\end{array}\\
%\begin{array}{ccc}
%a_5+a_6, & a_1+a_3+a_5, & a_2+a_4+a_6.
%\end{array}
%\end{array}
%\nonumber
%\end{equation}
%Note in particular that there appear R-charges that are not sums of consecutive $a_i$. Moreover the trial central charge before a-maximization is not equal
% to that of the other models (it is equal  only if, for example, $a_1=a_4$). 
%This more intricate distribution of R-charges may be due to the high degree of symmetry 
%of the toric diagram of $dP_3$.
  
\subsection{Orbifolds and singular horizons} 
\label{singhorizon}

In this subsection we deal with the problem of toric cones over non smooth five dimensional horizons; their toric diagram is characterized by the fact that some of its sides
pass through integer points: let's call $p$ the total number of such points on the sides. 
The global symmetries are now $d+p-1$. So we have to add new variables to the $a_i$, $i=1,\ldots d$ if we want to find all the global charges. 
Let's call the new variables $b_i$, $i=1,\ldots p$. 

\begin{figure}
\begin{minipage}[t]{0.48\linewidth}
\vspace{0pt}
\centering
\includegraphics[scale=0.63]{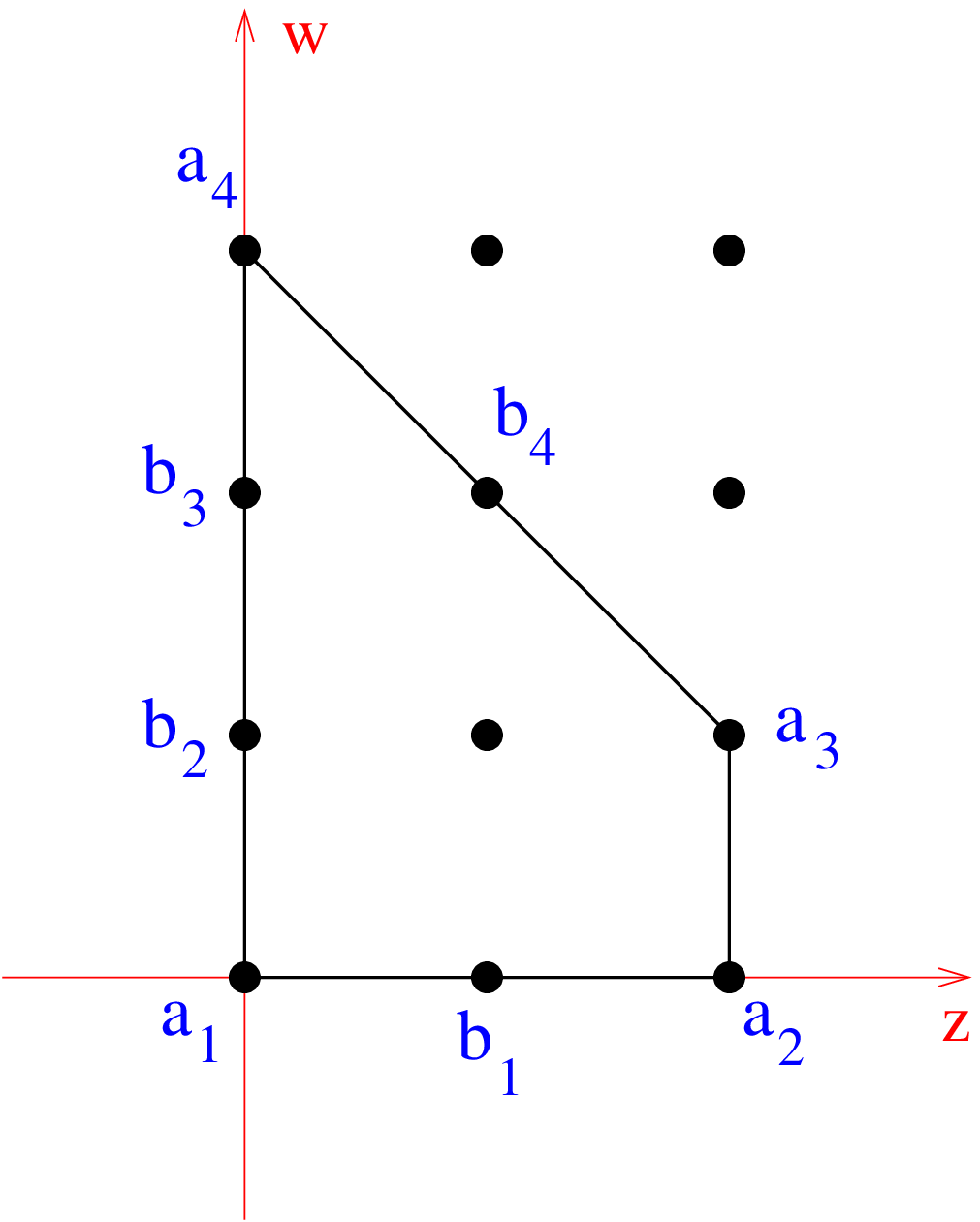}
\caption{Toric diagram for $L^{2,6;2,6}$.}
\label{toricl2626}
\end{minipage}%
~~~~~~\begin{minipage}[t]{0.48\linewidth}
\vspace{18pt}
\centering
\includegraphics[scale=0.56]{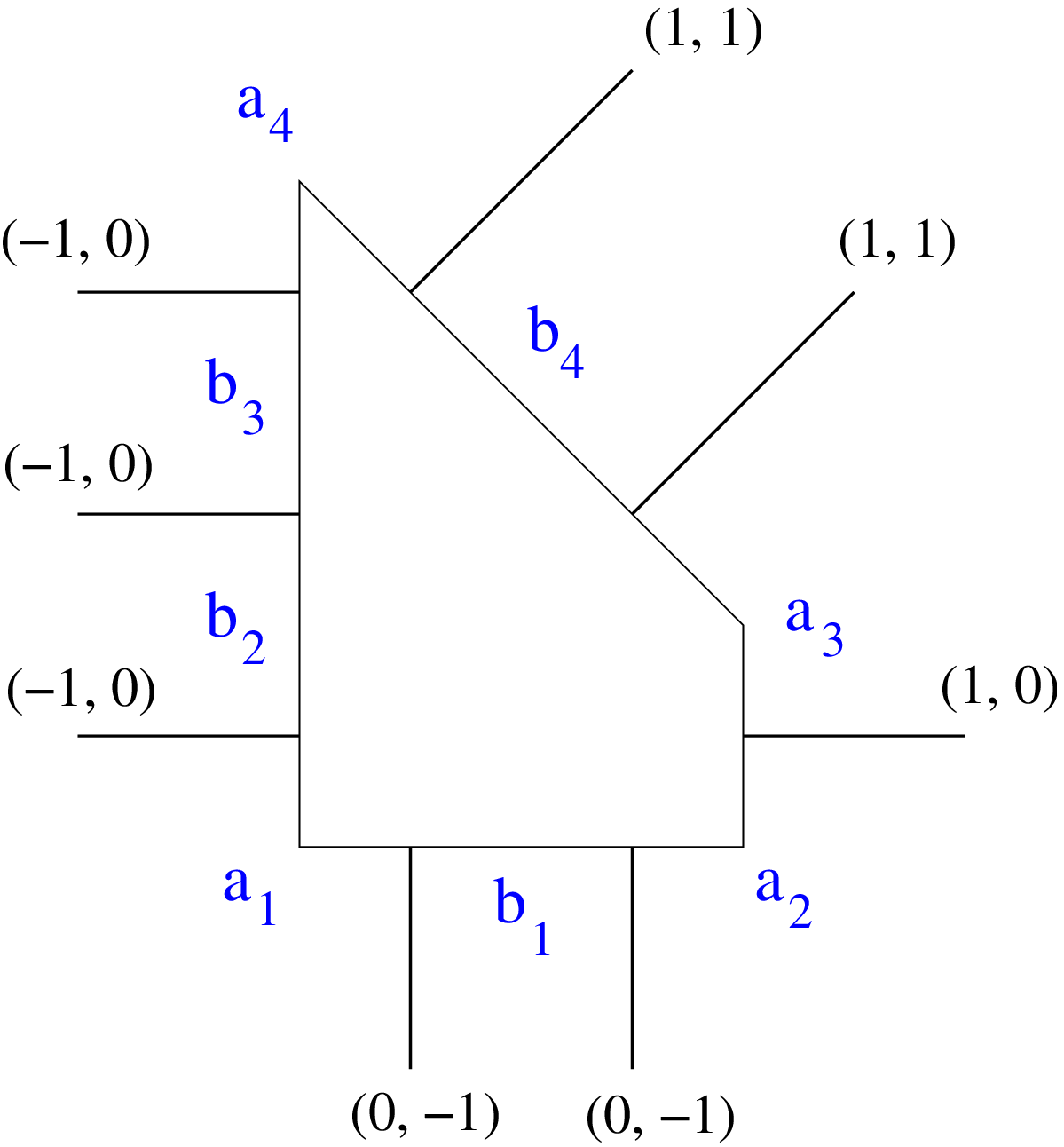}
\vspace{2pt}
\caption{The $(p,q)$ web for $L^{2,6;2,6}$.}
\label{pqwebl2626}
\end{minipage}%
%\end{figure}
%\vspace{0pt}
%\begin{figure}

\vspace{2pt}
\begin{minipage}[t]{0.48\linewidth}
\vspace{0pt}
\centering
\includegraphics[scale=0.45]{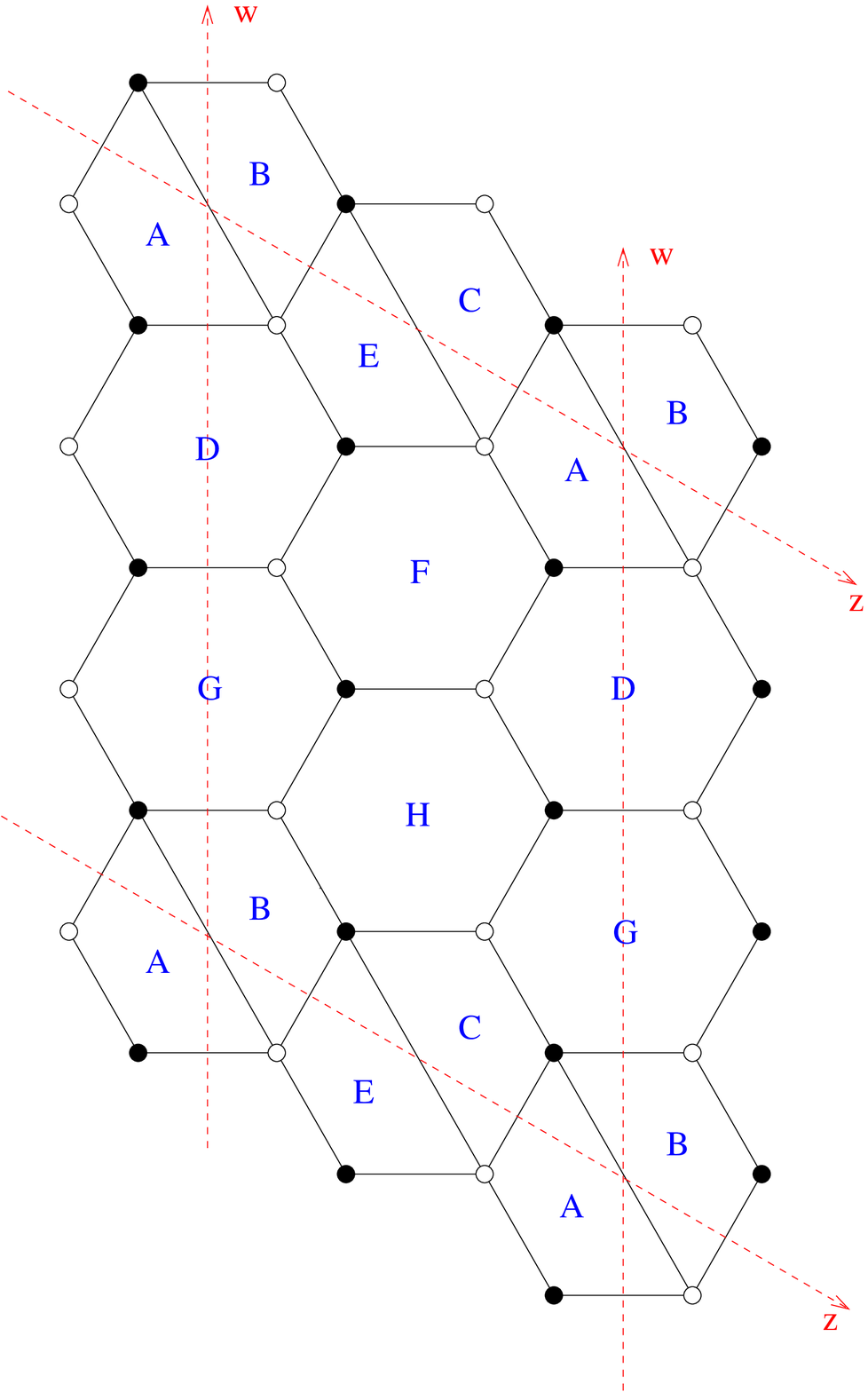}
\caption{Model I for $L^{2,6;2,6}$.}
\label{notsmooth1}
\end{minipage}%
~~~~~~\begin{minipage}[t]{0.48\linewidth}
\vspace{0pt}
\centering
\includegraphics[scale=0.45]{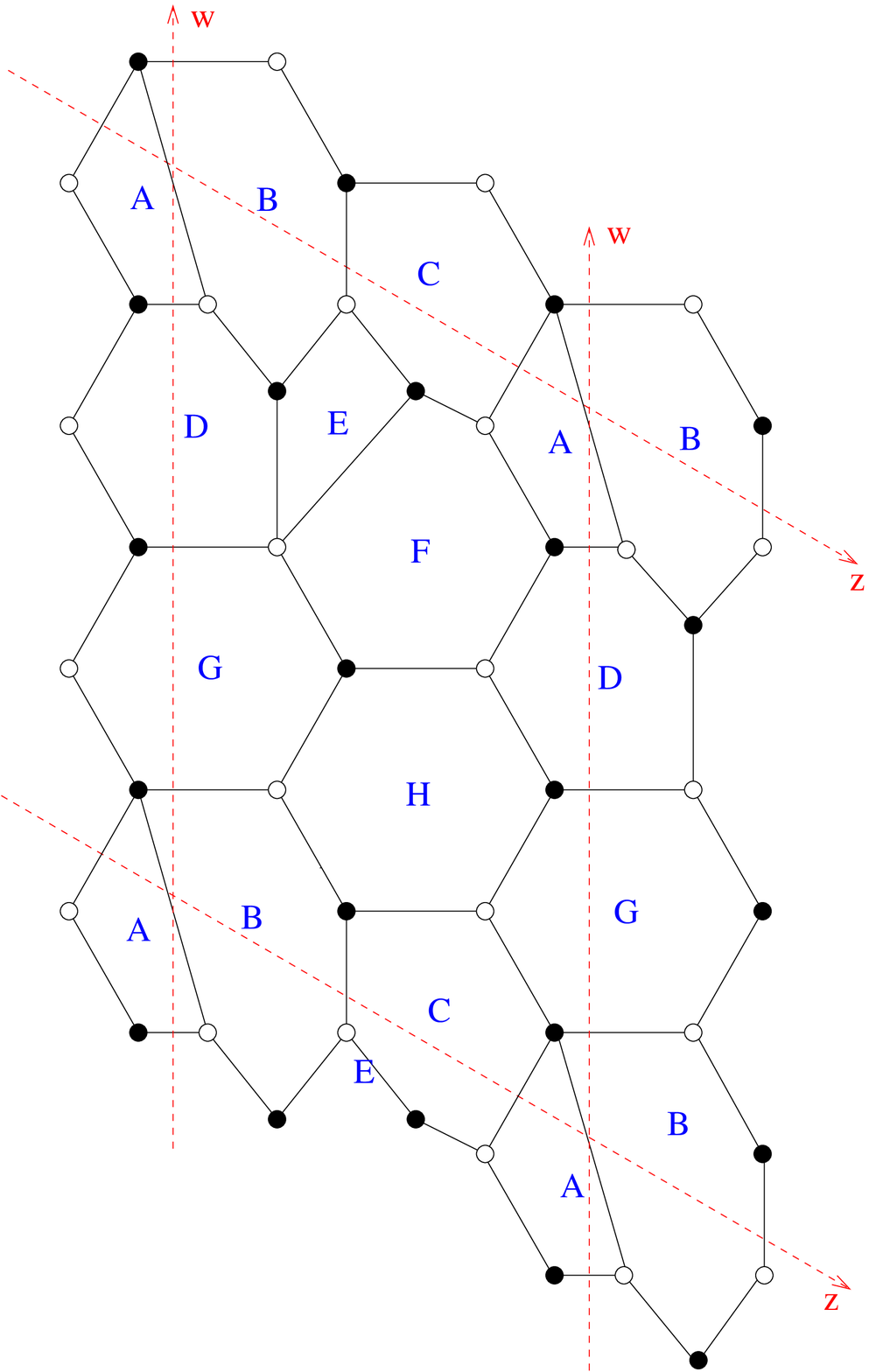}
\caption{Model II for $L^{2,6;2,6}$.}
\label{notsmooth2}
\end{minipage}%
\end{figure}

For simplicity we shall work on a specific example: a particular realization of $L^{2,6;2,6}$ whose toric diagram and $(p,q)$ web are drawn in Figures \ref{toricl2626}
and \ref{pqwebl2626}. This example has $d=4$
and $p=4$. The double area of the toric diagram is $F=8$

We have considered two toric phases of $L^{2,6;2,6}$. Their dimers are represented in Figures \ref{notsmooth1} and \ref{notsmooth2} respectively.
In fact it is not difficult to get the gauge theory by partial resolution of $C^3/(\mathbb{Z}_3 \times \mathbb{Z}_3)$, by resolving the point of coordinates $(3,0)$. The orbifold has 9 gauge groups and its gauge theory is described in \cite{dimers}. The only way to get a theory with 8 gauge groups is by eliminating
 (any) one of the links in the dimer of $C^3/(\mathbb{Z}_3 \times \mathbb{Z}_3)$. Integrating out the massive fields one gets Model II, Figure \ref{notsmooth2}, which has 
$E=22$, $V=14$. Performing a Seiberg duality with respect to the gauge group corresponding to face E in the dimer of Figure \ref{notsmooth2}, one gets Model I for this 
theory, which has fewer fields: $E=20$, $V=12$.   

\begin{figure}
\centering
\includegraphics[scale=0.5]{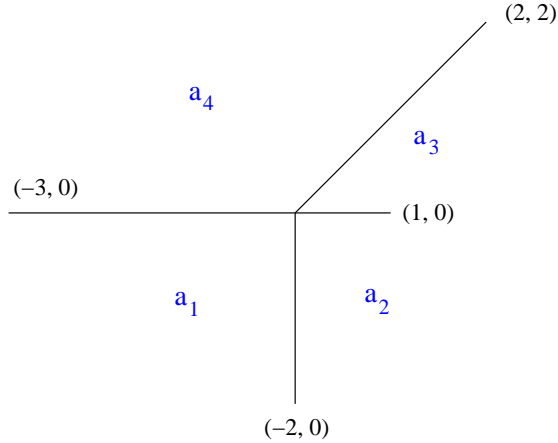}
\caption{A rough version of the (p,q) web for $L^{2,6;2,6}$.}
\label{pqweba}
\end{figure}

We identify Model I with the toric phase with a minimal number of fields for which our formulae should work. In fact it is possible to extend the algorithm described in Section \ref{geometry} to extract multiplicities from the toric diagram. Now one should assign charge $a_i$ to the $d$ vertices $V_i$ and $b_j$ to the $p$ integer points along the edges of $P$. Then the multiplicities are extracted using all the vectors of the $(p,q)$ web as in Figure \ref{pqwebl2626}. In our particular example one gets the fields:
\begin{equation}
\begin{array}{l}
\begin{array}{ccccccc}
a_1, & a_1+b_1, & b_1+a_2, & b_1+a_2+a_3, & b_1+a_2+a_3+b_4, & a_2, & a_2+a_3, 
\end{array}\\
\begin{array}{ccccccc}
a_2+a_3+b_4, & a_3, & a_3+b_4, &b_4+a_4, & b_4+a_4+b_3, & b_4+a_4+b_3+b_2, & a_4,
\end{array}\\
\begin{array}{cccccc}
 a_4+b_3, & a_4+b_3+b_2, & b_3+b_2+a_1, & b_3+b_2+a_1+b_1, & b_2+a_1, & b_2+a_1+b_1 
\end{array}
\end{array}
\label{molti}
\end{equation}
all with multiplicity equal to one (the total number of fields is thus 20, as in Model I). Note that, differently from the case of $a_i$, there is no chiral field 
with charge, say, $b_1$, since the $b_i$ are always included between parallel vectors (forming a parallelogram with area zero). Indeed it is not difficult to find a distribution of R-charges in the dimer configuration of Model I with these kinds of fields. Remember that the constraints are:
\begin{equation}
\sum_{i=1}^d a_i+ \sum_{j=1}^p b_j =2
\end{equation}
if we are dealing with R-charges, and
\begin{equation}
\sum_{i=1}^d a_i+ \sum_{j=1}^p b_j =0
\end{equation}
if we are dealing with global charges.
The trial R-charge depends both on $a_i$ and $b_j$, however we have verified in this case that the point that maximizes the central charge has all $b_i$ equal to zero. We conjecture that this may be true in general. In practice one could have started with the $(p,q)$ web drawn in Figure \ref{pqweba} for $L^{2,6;2,6}$; this is simply built ignoring the fact that there are points on the sides of $P$: the vectors are not the primitive ones, but they have the same length as the vectors of $P$.
Using the usual method for multiplicities as in Section \ref{geometry} with the $(p,q)$ web in Figure \ref{pqweba}, we get this table of multiplicities 
for the 20 fields:
\begin{equation}
\begin{array}{l@{\hspace{2em}}c@{\hspace{1em}}c@{\hspace{1em}}c@{\hspace{1em}}cc}
\mathrm{R-charge:} & a_1 & a_2 & a_3 & a_4 & a_2+a_3\\[0.5em] 
\mathrm{multiplicity:} & 6 & 2 & 2 & 6 & 4
\end{array}
\label{zerob}
\end{equation}   
to which (\ref{molti}) obviously reduces after setting $b_i=0$. Then the a-maximization can be performed also keeping into account only the charges $a_i$ and it is easy
 to check in this example that it reproduces the volumes of Z-minimization.

Let make also some comments about the generalization of the method described in subsection \ref{distribution} for assigning (R-)charges. The multiplicities of perfect matchings associated with vertices are again equal to one. Then we assign to the corresponding perfect matching (R-)charge $a_i$. But in general there is more than one perfect matching corresponding to a certain point along a side of $P$. In Model I of the example at hand the multiplicities of perfect matchings corresponding to
 points $b_1$, $b_2$, $b_3$, $b_4$ are respectively 2, 3, 3, 2. Therefore for every point along the sides we can choose a particular perfect matching and give it (R-)charge
 $b_i$ (and zero charge to all other perfect matchings). Then we can compute the charge of chiral fields as sums of charges of the perfect matchings to which they belong, as in subsection \ref{distribution}. In this way one always find R-charges (or global charges). However not all charges built in this way are linearly independent: this depends on the choice of perfect matchings. We verified in the case at hand that there are choices of perfect matchings for the $b_i$ that allow to find all the 7 independent (R-)charges,
some of them also reproducing the fields content given in (\ref{molti}).

The same conclusions hold for Model II of $L^{2,6:2,6}$. The number of fields now is 22: again a-maximization can be performed by setting to zero the $b_i$. We have all the fields appearing in table \ref{zerob} plus one field with charge $a_1+a_2$ and one with charge $a_3+a_4$, so that the trial R-charge is the same as in Model I for the mechanism described at the end of Section \ref{geometry}.

In conclusion in this subsection we have generalized our results to the case of non smooth horizon, checking in detail the algorithms on a particular example. This analysis deserves further study in order to verify whether it is true in the general case. In particular we guess that charges associated with points along the sides of the toric diagram are never relevant for a-maximization.
  
\section{Conclusions}
In this paper we computed the central charge and the R-charges of chiral fields
for all the superconformal gauge theories living on branes at toric
conical singularities. We also showed that the a-maximization technique
\cite{intriligator} is completely equivalent to the volume minimization
technique proposed in \cite{MSY}. This, by itself, is an absolutely
general check of the AdS/CFT correspondence, valid for all toric singularities.
 
In this general construction, something is obviously missing. We have now,
using the tiling construction \cite{dimers}, a direct determination of
the singularity associated with a given gauge theory. The inverse
process is still incomplete: we can determine R-charges and multiplicities
of fields but not the specific distribution of bi-fundamentals in
the quiver theory. We are quite confident that, in the long period, 
the dimers technology will allow to define a one-to-one correspondence 
between CFTs and toric singularities.
 
It would be also interesting to derive the assignment of charges and 
multiplicities we propose here. A possible way of deriving it goes
though mirror symmetry. It would be interesting to perform the analysis
done in \cite{hananymirror} in the general case. This analysis would
probably teach us also about the many toric phases that are associated with
the same superconformal gauge theory. 
\vskip 1truecm

\noindent {\Large{\bf Acknowledgments}}

\vspace{0.5em}

This work is supported in part by by INFN and MURST under 
contract 2001-025492, and by 
the European Commission TMR program HPRN-CT-2000-00131.

\vskip 2truecm

\noindent
{\Large{\bf Appendix}}
\renewcommand{\theequation}{A.\arabic{equation}}
\renewcommand{\thesubsection}{A.\arabic{subsection}}
\setcounter{equation}{0}\setcounter{section}{0}
\vskip 0.2truecm
\noindent

\subsection{A useful formula}
\label{useful}

Let us define the sets $C_h$, $h=1,2\ldots d$ which are subsets of $C$: a couple $(i,j)$ is in $C_h$ iff the R-charge of the corresponding chiral field is a sum $a_{i+1}+a_{i+2}+ \ldots a_{j}$ containing $a_h$. In practice $C_h$ is made up of all the couples $(i,j)$ such that the region $a_h$ in Figure \ref{pqweb} is contained in the angle $\leq 180^o$ generated by $v_i$ and $v_j$.

In this Appendix we shall prove the useful formula
\begin{equation}
S_h \equiv \sum_{(i,j) \in C_h} |\langle v_i, v_j \rangle|= \frac{V}{2} 
\label{sommah}
\end{equation} 
where $V$ is defined as: 
\begin{equation}
V \equiv E-F
\end{equation}
$V$ is the number of vertices of the associated dimer configuration. Note that (when the convex polygon $P$ has integer coordinates) equation (\ref{sommah}) proves also that $V$ is even. This agrees with the fact that there is an equal number of white and black vertices in the dimer configuration.

\begin{figure}
\begin{minipage}[t]{0.48\linewidth}
\centering
\includegraphics[scale=0.65]{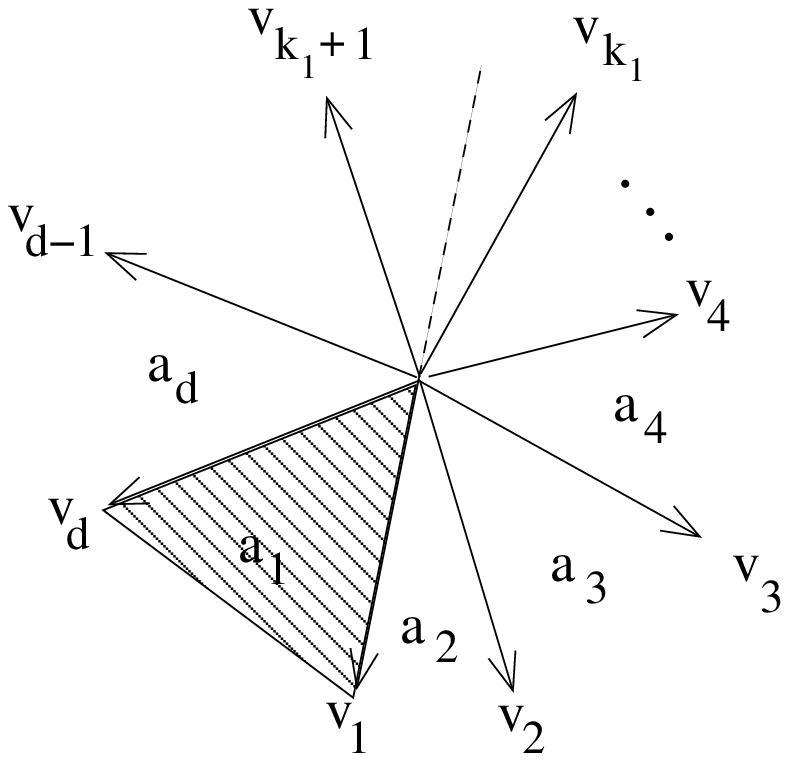}
\caption{Summing multiplicities in $C_1$.}
\label{app1}
\end{minipage}%
~~~~~~\begin{minipage}[t]{0.48\linewidth}
\centering
\includegraphics[scale=0.65]{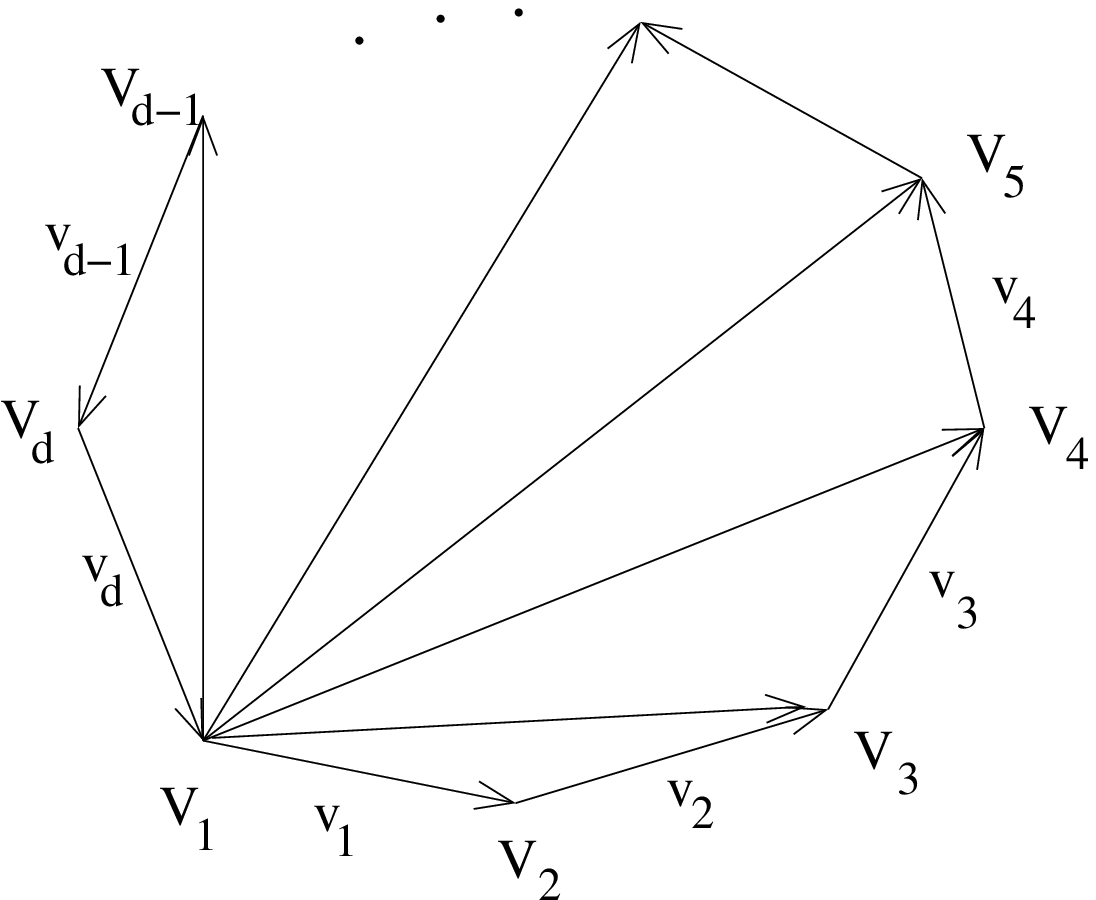}
\caption{Computing the area of $P$.}
\label{app2}
\end{minipage}%
\end{figure}

Given a vector $v_j$ in the $(p,q)$ web let us extend it (as in Figure \ref{app1} for the case $j=1$) and call $v_{k_j}$ the vector in the $(p,q)$ web just before this extension (moving in counter-clockwise direction). Note that:
\begin{equation} 
\begin{array}{l}
|\langle v_{j}, v_{j+1} \rangle|+|\langle v_{j}, v_{j+2} \rangle| \ldots +|\langle v_{j}, v_{k_j} \rangle|
-|\langle v_{j}, v_{j-1} \rangle|-|\langle v_{j}, v_{j-2} \rangle| \ldots -|\langle v_{j}, v_{k_j+1} \rangle| \\[0.5em]
= \langle v_{j}, v_{j+1}+ v_{j+2}+\ldots  v_{k_j} + v_{k_j+1}+ \ldots v_{j-1} \rangle \\[0.5em]  = -\langle v_j, v_j \rangle =0
\end{array}
\label{diff}
\end{equation}
where we have used that the sum of all $v_i$ in the $(p,q)$ web is zero. Remember that our indexes are always defined modulo $d$.
Note that equation (\ref{diff}) is just the difference $S_{j+1}-S_j$, so we have proved that all $S_j$ are equal.

To prove (\ref{sommah}) we can choose $h=1$ by a relabeling of vertices and sides (see Figure \ref{app1}). Let us consider the vector $v_1$ and write in the first line of a
 table all the multiplicities made up with $v_1$ (see below). We divide this line into two parts: on the left we write the pairs from $|\langle v_1,v_2 \rangle|$ to
 $|\langle v_1,v_{k_1} \rangle|$ (those which do not contain $a_1$) and on the right the pairs from $|\langle v_1,v_d \rangle|$ to $|\langle v_1,v_{k_1+1} \rangle|$ (that contain $a_1$)\footnote{If there is a vector $v_j$ lying just on the extension of $v_1$, the multiplicity $|\langle v_1, v_j\rangle|$ is zero and so it can be ignored.}. We repeat this procedure writing in the second line of the table all the pairs in $C$ that contain $v_2$, again dividing the line into two parts: on the left the pairs from $|\langle v_2,v_3 \rangle|$ to $|\langle v_2,v_{k_2} \rangle|$ and on the right the pairs from $|\langle v_2,v_1 \rangle|$ to
 $|\langle v_2,v_{k_2+1} \rangle|$.
We continue to fill in the lines with this ordering up to line $k_1$; in the remaining lines from $k_1+1$ to $d$ we reverse the order in which we divide lines in a left and right part. For example line $k_1+1$ contains the multiplicities formed with $v_{k_1+1}$ and we write on the left the pairs from $|\langle v_{k_1+1},v_{k_1} \rangle|$ to 
$|\langle v_{k_1+1},v_{k_{k_1+1}+1} \rangle|$ and on the right the pairs from $|\langle v_{k_1+1},v_{k_1+2} \rangle|$ to $|\langle v_{k_1+1},v_{k_{k_1+1}} \rangle|$: the idea is that all the pairs on the left do not contain $a_1$ whereas the pairs on the right may or may not contain $a_1$.

\[
%\begin{small}
\verb|  | \hspace{-4em}%
\begin{array}{ll@{\hspace{-0.1em}}ll@{\hspace{3em}}ll@{\hspace{-0.1em}}ll}
|\langle v_1,v_2 \rangle| & |\langle v_1,v_3 \rangle| & \ldots & |\langle v_1,v_{k_1} \rangle| & |\langle v_1,v_d \rangle| & |\langle v_1,v_{d-1} \rangle| & \ldots &
 |\langle v_1,v_{k_1+1} \rangle| \\[0.4em]
|\langle v_2,v_3 \rangle| & |\langle v_2,v_4 \rangle| & \ldots & |\langle v_2,v_{k_2} \rangle| & |\langle v_2,v_1 \rangle| & |\langle v_2,v_d \rangle| & \ldots & 
 |\langle v_2,v_{k_2+1} \rangle| \\
\rule{2em}{0pt} \vdots & & & & \rule{2em}{0pt}\vdots & & & \\
|\langle v_{k_1},v_{k_1+1} \rangle| & |\langle v_{k_1},v_{k_1+2} \rangle| & \ldots & |\langle v_{k_1},v_{k_{k_1}} \rangle| & |\langle v_{k_1},v_{k_1-1} \rangle| &
 |\langle v_{k_1},v_{k_1-2} \rangle| & \ldots & |\langle v_{k_1},v_{k_{k_1}+1} \rangle| \\[0.4em]
|\langle v_{k_1+1},v_{k_1} \rangle| & |\langle v_{k_1+1},v_{k_1-1} \rangle| & \ldots & |\langle v_{k_1+1},v_{\ldots} \rangle| & |\langle v_{k_1+1},v_{k_1+2} \rangle| &
 |\langle v_{k_1+1},v_{k_1+3} \rangle| & \ldots & |\langle v_{k_1+1},v_{\ldots} \rangle| \\
\rule{2em}{0pt}\vdots & & & & \rule{2em}{0pt}\vdots & & & \\
|\langle v_d,v_{d-1} \rangle| & |\langle v_d,v_{d-2} \rangle| & \ldots & |\langle v_d,v_{k_d+1} \rangle| & |\langle v_d,v_1 \rangle| & |\langle v_d,v_2 \rangle| & \ldots &
 |\langle v_d,v_{k_d} \rangle| \\
\end{array}
%\end{small}
\]
Note that the multiplicity associated with every pair of vectors $v_i$, $v_j$ in $C$ appears twice in the above table: once in line $i$ and once in line $j$. Hence the total sum of multiplicities in the table is $2 E$. But the sum of multiplicities in each line on the left equals the sum of multiplicities on the right in the same line because of (\ref{diff}). Hence the total sum of multiplicities on the right (or left) side of the table equals $E$. 
Moreover the pairs $(i,j)$ of vectors with multiplicity $|\langle v_i, v_j \rangle|$ in the left side of this table do not belong to $C_1$. All the pairs of vectors in $C_1$ appear (twice) in the right side of the table; but on the right side there appear also pairs that do not belong to $C_1$: $|\langle v_2, v_1 \rangle|$ in the second line, $|\langle v_3, v_2 \rangle|$ and $|\langle v_3, v_1 \rangle|$ in the second line, and so on. The total sum of such multiplicities is:
\begin{equation}
\begin{array}{l}
\langle v_1, v_2 \rangle + \langle v_1+v_2, v_3 \rangle \ldots + \langle v_1 + v_2 \ldots 
+ v_{k_1-1}, v_{k_1}  \rangle +\\
+ \langle v_{k_1+1}, v_{k_1+2} + v_{k_1+3} \ldots + v_d \rangle  \ldots + \langle v_{d-1}, v_d \rangle = \\[0.5em]
= \langle v_1, v_2 \rangle + \langle v_1+v_2, v_3 \rangle \ldots + \langle v_1 + v_2 \ldots 
+ v_{k_1-1}, v_{k_1}  \rangle + \\
\langle v_1 + v_2 \ldots + v_{k_1}, v_{k_1+1}  \rangle + \langle v_1 + v_2 \ldots + v_{d-2}, v_{d-1} \rangle = 
2 \mathrm{Area}(P)=F 
\end{array}
\end{equation}  
where in the first equality we have used that the sum of all $v_i$ is zero and the bilinearity and antisymmetry of the determinant. The sum of all multiplicities in the right side of the table above that do not belong to $C_1$ is thus equal to the double area of $P$, see Figure \ref{app2}.
The sum we had to compute is therefore: 
\begin{equation}
S_1 \equiv \sum_{(i,j) \in C_1} |\langle v_i, v_j \rangle|= \frac{E-F}{2}=\frac{V}{2} 
\end{equation}
which is relation (\ref{sommah}).

\subsection{Charges}
We now show that our proposed formula for extracting multiplicities of chiral fields from the toric diagram correctly gives $U(1)$ baryon, flavor and R-charges with trace equal to zero.
Let's start with a charge commuting with supersymmetry; as explained in Section \ref{geometry} it can be built by assigning charges $a_i$ to chiral fields associated with vectors $V_i$ of the fan with (\ref{sommazero}):
\begin{equation}
\sum_{i=1}^d a_i =0 
\label{sumglobal}
\end{equation}  
Therefore we have $d-1$ global symmetries, 2 of which are flavor symmetries and the remaining $d-3$ are baryonic symmetries (remember that for non smooth horizons we have to consider also charges associated to integer points lying along the sides of the convex polygon $P$; the total sum of all charges associated to ``fundamental'' fields (\ref{sumglobal}) must still be zero). The charge of a generic ``composite'' chiral field associated with the pair $(i,j) \in C$ is simply the sum 
$a_{i+1}+\ldots a_j$. 

The trace of a generic $U(1)$ global symmetry is thus:
\begin{equation}
\begin{array}{ll}
\mathrm{tr} \, U(1) & =  \displaystyle \sum_{(i,j) \in C} 
|\langle v_i,v_j \rangle| \,\,(a_{i+1}+a_{i+2}\ldots +a_{j}) \\[1.5em] 
 & \displaystyle = \sum_{h=1}^d a_h \sum_{(i,j)\in C_h} |\langle v_i,v_j \rangle|= 
 \displaystyle \frac{V}{2}\sum_{h=1}^d a_h =0
\end{array} 
\end{equation} 
where we have used that $S_h$ in (\ref{sommah}) does not depend on $h$.

Let us now turn to R-symmetry; to build the generic trial R-symmetry we have to associate a R-charge $a_i$ to the chiral fields corresponding to divisors $V_i$ (and also to fields corresponding to vertices along sides for non smooth horizons); the only difference with the global case is that now the sum must satisfy (\ref{sommadue}): 
\begin{equation}
\sum_{i=1}^d a_i =2 
\label{sumr}
\end{equation}
The trace of a generic $U(1)_R$ symmetry is now
\begin{equation}
\begin{array}{ll}
\mathrm{tr} \, U(1)_R & =  F + \displaystyle \sum_{(i,j) \in C} 
|\langle v_i,v_j \rangle| \,\,(a_{i+1}+a_{i+2}\ldots +a_{j}-1) \\[1.5em] 
 & \displaystyle = F + \sum_{h=1}^d a_h \sum_{(i,j)\in C_h} |\langle v_i,v_j \rangle| -\sum_{(i,j) \in C} 
|\langle v_i,v_j \rangle| \\[1.5em]
 & = F + \displaystyle \frac{V}{2} \left( \sum_{h=1}^d a_h \right) -E =0
\end{array} 
\label{tracer}
\end{equation} 
where we have used equation (\ref{sommah}). The term $F=2\, \mathrm{Area}(P)$ comes from gauginos, since we know that the double area gives the number of gauge groups. This also shows that for gauge theories dual to toric geometries the trial R-charge always reduces to $a=9/32 \, \mathrm{tr} \,R^3$.

Let us now prove that the trace of cubic t'Hooft anomaly and mixed cubic anomaly for baryonic symmetries are always zero with the multiplicities and charges for chiral fields that we have conjectured in this paper. The vanishing of such anomalies is required by the AdS/CFT correspondence and is always true for the quiver gauge theories under consideration since the global baryonic symmetries are (the non anomalous) linear combinations of the $U(1)$ part of the original gauge groups $U(N)$ (after the AdS/CFT limit they generally become $SU(N)$ gauge groups). But since we have only conjectured the multiplicities of chiral fields and a full algorithm for extracting the whole gauge theory from toric geometry is still lacking, the proof of zero cubic anomaly for baryonic symmetries is a non trivial check of our conjecture.

First of all recall that, as discovered in \cite{tomorrow}, the $d-1$ baryonic symmetries are simply the linear relations between the $d$ generators of the toric fan $V_i$: 
\footnote{Again recall that for non smooth horizons one has to add to the set of $V_i$ all the vectors in the fan arriving at the integer points along the sides of $P$.} 
if $(a_1,a_2,\ldots a_d)$ are the charges of a baryonic symmetry associated with chiral fields corresponding to the vectors $V_i$ we have equation (\ref{linear})
\begin{equation}
\sum_{i=1}^d  a_i V_i=0
\end{equation}
Knowing that $V_i$ have first coordinate equal to 1, and that the other two components are the coordinates $(x_i,y_i)$ of the vertices of $P$ in the plane, the previous equation
 can also be restated by saying that $(a_1,a_2,\ldots a_d)$ must satisfy (\ref{sumglobal}), as all global symmetries, and moreover the constraint:
\begin{equation}
a_2 v_1+a_3 (v_1+v_2) \ldots + a_d(v_1+v_2+\ldots v_{d-1})=0 
\label{relations}
\end{equation}  
where we have started to compute the coordinates of the vertices of $P$ from the first vertex (see Figure \ref{app2}), but one could have started from any other point in the plane of $P$ because of (\ref{sumglobal}).
Note also that a basis for the two flavor symmetries orthogonal to the baryonic ones is given by the $x$ and $y$ coordinates of the vertices of $P$ 
in the plane containing $P$ referred to the barycenter of $P$, so that (\ref{sumglobal}) holds. 

So take now three different (or equal) baryonic symmetries: $(a_1,a_2,\ldots a_d)$, $(a'_1,a'_2,$ $\ldots a'_d)$  and  $(b_1,b_2,\ldots b_d)$ all satisfying (\ref{sumglobal}) and (\ref{relations}). To avoid writing too long formulae we will consider first the case when two symmetries are equal, say $a_i=a'_i$, and then we will extend our results to the general case. The mixed cubic t'Hooft anomaly with our formula for multiplicities becomes:
\begin{eqnarray}
\mathrm{tr} \, \left( U(1)_B^a \right)^2 \, U(1)_B^b & = &
\displaystyle \sum_{(i,j) \in C} |\langle v_i,v_j \rangle| \,(a_{i+1}+a_{i+2}\ldots + a_j)^2 (b_{i+1}+b_{i+2} \ldots b_j) \nonumber \\ 
& = & \displaystyle \sum_{h=1}^d b_h \left(\displaystyle \sum_{(i,j) \in C_h} |\langle v_i,v_j \rangle| \,(a_{i+1}+a_{i+2}\ldots + a_j)^2 \right) \nonumber\\
 & \equiv & \sum_{h=1}^d b_h c_h
 \label{cubic}
\end{eqnarray}
where the coefficients $c_h$ are defined by the last equality. We have to prove that the vector formed by $c_h$ is orthogonal to a generic baryonic symmetry, that is that the vector 
of $c_h$ is a linear combination of $x$ and $y$ coordinates of vertices of $P$ up to some multiple of $(1,\ldots,1)$.
So let's compute the differences:
\begin{eqnarray}
& & c_{j+1}-c_{j} = \nonumber \\ 
 & & = |\langle v_j, v_{j+1} \rangle| \, \left( a_{j+1} \right)^2 + |\langle v_j, v_{j+2} \rangle| \, \left( a_{j+1}+a_{j+2} \right)^2 \ldots
 + |\langle v_j, v_{k_j} \rangle| \, \left( a_{j+1}+a_{j+2}\ldots +a_{k_j} \right)^2 \nonumber \\
 & & -|\langle v_j, v_{j-1} \rangle| \, \left( a_j \right)^2 - |\langle v_j, v_{j-2} \rangle| \, \left( a_j+a_{j-1} \right)^2 \ldots 
 - |\langle v_j, v_{k_j+1} \rangle| \, \left( a_j+a_{j-1}\ldots + a_{k_j+2} \right)^2 \nonumber
\end{eqnarray}
where there survive only the sum over pairs that contain $a_{j+1}$ and do not contain $a_j$, minus the sum over pairs that contain $a_j$ and do not contain $a_{j+1}$, since all other pairs cancel. The symbols $k_j$ are defined as in Appendix \ref{useful}. The previous equation can be rewritten as
\begin{equation}
c_{j+1}-c_j=\langle v_j , T_j \rangle
\label{diffc}
\end{equation}
where $T_j$ is the vector:
\begin{eqnarray}
T_j & = & v_{j+1} \left( a_{j+1} \right)^2+v_{j+2} \left( a_{j+1}+a_{j+2} \right)^2 \ldots + v_{k_j} \left( a_{j+1} + a_{j+2} \ldots+ a_{k_j} \right)^2 \nonumber \\
 & & + v_{j-1} \left( a_j \right)^2+v_{j-2} \left( a_{j}+a_{j-1} \right)^2 \ldots + v_{k_j+1} \left( a_j + a_{j-1} \ldots + a_{k_j+2} \right)^2 \nonumber \\
 & = &  v_{j+1} \left( a_{j+1} \right)^2+v_{j+2} \left( a_{j+1}+a_{j+2} \right)^2 \ldots + v_{k_j} \left( a_{j+1} + a_{j+2} \ldots + a_{k_j} \right)^2 \nonumber \\
 & & + v_{k_j+1} \left( a_{j+1} + a_{j+2} \ldots a_{k_j+1} \right)^2  \ldots % + v_{j-2}\left( a_{j+1} + a_{j+2} \ldots a_{j-2} \right)^2
    + v_{j-1}\left( a_{j+1} + a_{j+2} \ldots + a_{j-1} \right)^2 
\end{eqnarray}
where in the last line we have reordered the sum and used that the sum of all $a_i$ is zero (\ref{sumglobal}).

Now we want to show that all vectors $T_j$ are equal: $T_1=T_2 \ldots=T_d \equiv T$; it is enough to prove that consecutive vectors $T_j$ are equal and, by a relabeling of vectors and vertices, it is enough to prove this for, say $T_1$ and $T_2$. A straightforward computation then yields:
\begin{eqnarray}
 & & T_2-T_1 = \nonumber \\[0.2em]
 & = &  v_3 \left( a_3 \right)^2 + v_4 \left( a_3 +a_4 \right)^2 \ldots + v_d \left( a_3 +a_4\ldots + a_d \right)^2 
       + v_1 \left( a_3 +a_4\ldots + a_d+a_1 \right)^2 \nonumber \\
 & & - v_2 \left( a_2 \right)^2 - v_3 \left( a_2 + a_3 \right)^2 -  v_4 \left( a_2 + a_3 + a_4 \right)^2 \ldots - v_d \left(a_2+a_3 \ldots +a_d \right)^2 \nonumber \\[0.2em]
 & =  & -a_2^2 \left( v_2+v_3 \ldots + v_d \right) - 2 a_2 \left[v_3 a_3 + v_4 \left( a_3+a_4 \right)\ldots + v_d \left(a_3+a_4 \ldots a_d \right) \right] 
        + v_1 \left( a_2 \right)^2 \nonumber \\[0.2em]
 & = & 2 a_2^2 \, v_1 - 2 a_2  \left[ v_3 a_3 + v_4 \left( a_3+a_4 \right) \ldots + v_d \left(a_3+a_4 \ldots a_d \right) \right] \nonumber \\[0.2em]
 & = & -2 a_2 \left[ v_3 a_3 + v_4 \left( a_3+a_4 \right) \ldots + v_d \left( a_3+a_4 \ldots a_d \right)
        + v_1 \left( a_3+a_4 \ldots +a_d+a_1 \right)\right] \nonumber \\[0.2em]
 & = & -2 a_2 \left[ a_1 v_1 + a_d \left( v_1+v_d  \right) + a_{d-1} \left( v_1+v_d+v_{d-1} \right) \ldots 
              + a_3 \left( v_1+v_d+v_{d-1} \ldots +v_3 \right)  \right] \nonumber \\[0.2em]
 & = & 0
\label{t2t1} 
\end{eqnarray} 
where we have used (\ref{sumglobal}) and that the sum of $v_i$ is zero. In the last step we have used that $(a_1,\ldots a_d)$ is a baryonic symmetry, since the last sum is one of the kind of (\ref{relations}), centered in the second vertex of the polygon $P$.

Now we get for the differences:
\begin{eqnarray}
c_2-c_1 & = & \langle v_1, T \rangle \nonumber \\
c_3-c_1 & = & \left( c_3-c_2 \right) + \left( c_2 - c_1 \right)= \langle v_1 + v_2, T \rangle  \nonumber\\
\vdots & & \nonumber \\
c_d-c_1 & = & \langle v_1 + v_2 \ldots + v_{d-1}, T \rangle
\end{eqnarray} 
and for the cubic t'Hooft anomaly of baryonic symmetries:
\begin{eqnarray}
& & \mathrm{tr} \, \left( U(1)_B^a \right)^2 \, U(1)_B^b  =  \sum_{h=1}^d b_h c_h \nonumber \\[0.2em]
& & = c_1 \left( b_1+b_2+b_3 \ldots + b_d \right) + b_2 \left( c_2-c_1 \right)+ b_3 \left(c_3-c_1 \right) \ldots + b_d \left( c_d-c_1 \right) \nonumber \\[0.2em]
& & = \langle b_2 v_1 + b_3 \left( v_1+v_2 \right) \ldots + b_d \left(v_1+v_2 \ldots +v_{d-1} \right),T \rangle \nonumber\\[0.2em]
& & = 0 
\label{simplybaryon}
\end{eqnarray}
where we have used that $(b_1,\ldots b_d)$ is a baryonic symmetry thus satisfying (\ref{sumglobal}) and (\ref{relations}). It is easy to generalize to the case $a_i \neq a'_i$: the coefficients $c_h$ are given now by:
\begin{equation}
c_h = \displaystyle \sum_{(i,j) \in C_h} |\langle v_i,v_j \rangle| \, \left( a_{i+1}+a_{i+2}\ldots + a_j \right)  
 \left( a'_{i+1}+a'_{i+2}\ldots + a'_j \right)
\end{equation}
and one has to repeat all the steps leading to (\ref{t2t1}) keeping products of sums of $a_i$ and $a'_i$ instead of squares. It is to see that now (\ref{t2t1}) reads:
\begin{eqnarray}
& & T_2-T_1 = \nonumber \\[0.2em]
& = & - a_2 \left[ a'_1 v_1 + a'_d \left( v_1+v_d  \right) + a'_{d-1} \left( v_1+v_d+v_{d-1} \right) \ldots 
              + a'_3 \left( v_1+v_d+v_{d-1} \ldots +v_3 \right)  \right] \nonumber \\
&  & - a'_2 \left[ a_1 v_1 + a_d \left( v_1+v_d  \right) + a_{d-1} \left( v_1+v_d+v_{d-1} \right) \ldots 
              + a_3 \left( v_1+v_d+v_{d-1} \ldots +v_3 \right)  \right] \nonumber \\[0.2em]
& = & 0                            
\end{eqnarray}
so that one has to use that both $a_i$ and $a'_i$ are baryonic. The proof then proceeds as before (\ref{simplybaryon}). 
This concludes our proof for the cubic anomaly of baryonic symmetries:
\begin{equation}
\mathrm{tr} \,  U(1)_B^a \, U(1)_B^{a'} \, U(1)_B^b = 0.
\end{equation}

\subsection{Decoupling baryon charges in a-maximization}

In this Appendix we shall prove equation (\ref{derbaryon}):
\begin{equation}
\sum_{h=1}^d b_h \frac{\partial a}{\partial a_h}_{|a_i=f_i(x,y)}=0
\label{decoupling}
\end{equation}
for every baryonic symmetry with charges $b_i$ for the chiral fields associated to $V_i$.
The functions $f_i(x,y)$ and $l_i(x,y)$ are defined as in (\ref{Rc}) and (\ref{li}):
\begin{equation}
f_i= \frac{2 \, l_i}{S}, \qquad l_i=\frac{\langle v_{i-1},v_i \rangle}{A_{i-1}\,A_i} 
\label{replace}
\end{equation}
where we have defined the sum
\begin{equation}
S \equiv \sum_{i=1}^d l_i  
\label{esse}
\end{equation} 
and the double area of triangles in Figure \ref{reeb}:
\begin{equation}
A_i\equiv \langle r_i, v_i \rangle = \langle r_{i+1}, v_i \rangle \, .
\end{equation}
Remember that $r_{i+1}-r_i=v_i$. Note that $l_i$ is positive inside the interior of $P$ and diverges on the edges $v_i$ and $v_{i-1}$.

We will need some useful relations among these quantities. In particular we can prove the vectorial identity:
\begin{equation}
l_i \, r_i = \frac{v_{i-1}}{A_{i-1}}-\frac{v_i}{A_i} \, . 
\label{split}
\end{equation}
In fact a straightforward computation gives
\begin{equation}
l_i \, r_i - \left( \frac{v_{i-1}}{A_{i-1}}-\frac{v_i}{A_i} \right)=
\frac{\langle v_{i-1},v_i \rangle r_i - v_{i-1} \langle r_i,v_i \rangle
 + v_i \langle r_i,v_{i-1} \rangle}{A_{i-1} \, A_i} \equiv \frac{N}{A_{i-1} \, A_i}
\end{equation}
Then we get for the numerator $\langle N, v_i \rangle=\langle N, v_{i-1} \rangle = 0$. So $N$ has to be parallel both to $v_i$ and $v_{i-1}$ which are two linearly independent vectors. Therefore $N=0$ and we have proved (\ref{split}).

By summing up (\ref{split}) we get another important property:
\begin{equation}
\sum_{i=1}^d l_i r_i =0
\label{weights}
\end{equation}
which says that the $l_i$ are (proportional to) the weights that should be put on the vertices $V_i$ of $P$ to keep it in equilibrium if we want to suspend it by the internal point $B$.

Equation (\ref{decoupling}) then reads
\begin{equation}
\sum_{h=1}^d b_h \frac{\partial a}{\partial a_h}_{|a_i=f_i(x,y)}= \frac{27}{32}\sum_{h=1}^d b_h d_h
\label{bhdh}
\end{equation}
where we have defined
\begin{equation}
d_h=\left(\displaystyle \sum_{(i,j) \in C_h} |\langle v_i,v_j \rangle| 
\,(a_{i+1}+a_{i+2}\ldots + a_j-1)^2 \right)_{|a_i=f_i(x,y)}
\end{equation}
that comes from deriving (\ref{afield}) with respect to $a_h$.
Note that (\ref{bhdh}) is, up to a constant factor, equal to $\mathrm{tr} R^2 b$, with $R$ the trial R symmetry and $b$ the baryon charge. Note in fact the similarities with equation (\ref{cubic}): the main difference here being that we are dealing with R-symmetry, so the constraint on $a_i$ is (\ref{sommadue}), automatically implemented by the substitution (\ref{replace}).  

Again the idea is to compute the differences $d_{j+1}-d_{j}$ and, similarly to (\ref{diffc}), to rewrite them as
\begin{equation}
d_{j+1}-d_{j}=\langle v_j, \tilde W_j \rangle  
\end{equation}
where now the vector $\tilde W_j$ reads:
\begin{equation}
\begin{array}{l}
\tilde W_j = \left[ v_{j+1} \left( a_{j+1}-1 \right)^2+v_{j+2} \left( a_{j+1}+a_{j+2}-1 \right)^2 \ldots + v_{k_j} \left( a_{j+1} + a_{j+2} \ldots+ a_{k_j} -1 \right)^2  \right. \nonumber \\
\left.  + v_{j-1} \left( a_j -1 \right)^2+v_{j-2} \left( a_{j}+a_{j-1} -1 \right)^2 \ldots + v_{k_j+1} \left( a_j + a_{j-1} \ldots + a_{k_j+2}-1 \right)^2\right]_{|a_i=f_i} \nonumber 
\end{array}  
\end{equation}  
where the symbols $k_j$ are defined as in Appendix \ref{useful}.
Performing the substitution (\ref{replace}) $a_i =f_i$ and taking the common denominator we get
\begin{eqnarray}
& & \hspace{-1.5em}  S^2 \, \tilde W_j = \nonumber \\
& = & v_{j+1} \left( l_{j+1}-l_{j+2}-l_{j+3}\ldots -l_j \right)^2+v_{j+2} \left( l_{j+1}+l_{j+2}-l_{j+3} \ldots - l_j           \right)^2 + \ldots \nonumber \\
& & + v_{k_j} \left( l_{j+1} + l_{j+2} \ldots+ l_{k_j} -l_{k_j+1}\ldots -l_j \right)^2 + \nonumber \\
& & + v_{j-1} \left( l_j - l_{j-1}-l_{j-2}\ldots -l_{j+1} \right)^2
    + v_{j-2} \left( l_{j}+l_{j-1} -l_{j-2}\ldots -l_{j+1} \right)^2 + \ldots \nonumber \\
& & + v_{k_j+1} \left( l_j + l_{j-1} \ldots + l_{k_j+2}-l_{k_j+1}\ldots -l_{j+1} \right)^2 \nonumber \\
& = & v_{j+1} \left( l_{j+1}-l_{j+2}-l_{j+3}\ldots -l_j \right)^2+v_{j+2} \left( l_{j+1}+l_{j+2}-l_{j+3} \ldots - l_j           \right)^2 + \ldots \nonumber \\
& & + v_{k_j} \left( l_{j+1} + l_{j+2} \ldots+ l_{k_j} -l_{k_j+1}\ldots -l_j \right)^2 + \nonumber \\
& & + v_{k_j+1} \left( l_{j+1} + l_{j+2} \ldots + l_{k_j+1}-l_{k_j+2}\ldots -l_j \right)^2 + \ldots \nonumber \\
& &   + v_{j-1} \left( l_{j+1} + l_{j+2} \ldots +l_{j-1} -l_j \right)^2
\end{eqnarray}
where in the last step we have reordered the sum. For later convenience, let us add to $\tilde W_j$ two terms proportional to $v_j$ defining the new vector $W_j$ as:
\begin{eqnarray}
& & \hspace{-1.5em}  S^2 \, W_j = \nonumber \\
& = & v_{j+1} \left( l_{j+1}-l_{j+2}-l_{j+3}\ldots -l_j \right)^2+v_{j+2} \left( l_{j+1}+l_{j+2}-l_{j+3} \ldots - l_j           \right)^2 + \ldots \nonumber \\
& &   + v_{j-1} \left( l_{j+1} + l_{j+2} \ldots +l_{j-1} -l_j \right)^2+ \\
& &   + v_j \left( l_{j+1} + l_{j+2} \ldots +l_{j-1} + l_j \right)^2 - 8 S \frac{v_j}{A_j}
\end{eqnarray}
and because of antisymmetry of the determinant we still have:
\begin{equation}
d_{j+1}-d_{j}=\langle v_j, W_j \rangle  
\end{equation}
We want to prove that all $W_j$ are equal: $W_1=W_2\ldots =W_d \equiv W$. As in the previous Appendix, it is enough 
to show the equality of consecutive $W_j$, $W_{j+1}$, and, up to a relabeling of indexes, it is enough to show that $W_2=W_1$. So let's compute the difference:
\begin{eqnarray}
& & \hspace{-1.5em} S^2 (W_2-W_1)= \nonumber \\
& = & v_{3} \left( l_{3}-l_{4}-l_{5}\ldots -l_1-l_2 \right)^2 + v_{4} \left( l_{3}+l_{4}-l_{5} \ldots - l_1-l_2              \right)^2 + \ldots \nonumber \\ 
& & + v_1 \left( l_3+l_4+l_5 \ldots +l_1-l_2  \right)^2 + v_2 \left( l_3+l_4+l_5 \ldots +l_1+l_2  \right)^2 \nonumber \\
& & - v_2 \left( l_2-l_3-l_4-l_5 \ldots -l_1 \right)^2 - v_3 \left( l_2+l_3-l_4-l_5 \ldots -l_1 \right)^2 \nonumber \\ 
& & -v_4 \left( l_2+l_3+l_4-l_5 \ldots -l_1 \right)^2 \ldots 
   - v_1 \left( l_2+l_3+\ldots +l_d+l_1 \right)^2 \nonumber \\
& & -8 S \left( \frac{v_2}{A_2}-\frac{v_1}{A_1} \right) \nonumber \\[0.5em]  
& = & 4 l_2 \left[ v_2(l_3+l_4 +l_5 \ldots +l_1) + v_3(-l_3+l_4 +l_5 \ldots +l_1) \right. \nonumber \\ 
& & \left. + v_4 (-l_3-l_4 +l_5 \ldots +l_1)\ldots +v_1 (-l_3-l_4 -l_5 \ldots -l_1) \right] \nonumber \\ 
& & +8 S \left( \frac{v_1}{A_1}-\frac{v_2}{A_2} \right)
\end{eqnarray}  
where in the last step we have computed the differences between factors with the same $v_i$ keeping in consideration that each time only the term $l_2$ changes relative sign. Now we reorder the first term in the square bracket and we use
equation (\ref{split}) (with $i=2$) for the last term:
\begin{eqnarray}
& & \hspace{-1.5em} S^2 (W_2-W_1)= \nonumber \\[0.5em]
& = & 4 l_2 \left[ l_3 \left( v_2-v_3-v_4 \ldots -v_1 \right) 
      + l_4  \left( v_2+v_3-v_4 \ldots -v_1 \right) +\ldots \right. \nonumber \\
&  & \left. + l_1  \left( v_2+v_3+v_4 \ldots + v_d -v_1 \right) \right] + 8 S l_2 r_2 \nonumber \\[0.5em]
& = & 8 l_2 \left[ l_3 v_2  + l_4  \left( v_2+v_3 \right)\ldots
      + l_1  \left( v_2+v_3 \ldots + v_d  \right) \right]  + 8 S l_2 r_2 \nonumber \\[0.5em]
& = & 8 l_2 \left[ l_2 r_2 + l_3 \left( r_2 + v_2 \right) + l_4 \left( r_2 + v_2 + v_3 \right)
      \ldots + l_1 \left( r_2 + v_2 + v_3\ldots + v_d \right) \right] \nonumber \\
& & - 8 l_2 r_2 \left(\sum_{j=1}^d l_j \right)  + 8 S l_2 r_2        
\end{eqnarray} 
where in the second equality we have used that $\sum_i v_i=0$, and in the third equality we have added and subtracted the same term. Now the last two terms cancel and, noting that $r_2+v_2+v_3\ldots v_{i-1}=r_i$ (look at Figure \ref{reeb}) the sum in the square brackets becomes:
\begin{equation}
S^2 (W_2-W_1)= 8 l_2 \left( \sum_{j=1}^d l_j r_j \right)=0
\end{equation}
where we have used (\ref{weights}). Hence we conclude that $W_1=W_2\ldots =W_d \equiv W$. Now the conclusion of the proof of (\ref{decoupling}), that is $\sum_h b_h d_h=0$, proceeds as in (\ref{simplybaryon}) (with the appropriate substitutions $T\rightarrow W$, $c_h \rightarrow d_h$). In this step we use that $b_i$ are baryonic. This concludes our proof.

\subsection{The equality of $a$ and $a^{MSY}$}
In this Appendix we give a general proof of equation (\ref{equal}), that shows the agreement of the central charge $a$ and the total volume even before maximization, once the substitution $a_i=f_i\equiv 2 l_i/S$ has been performed. 

Taking into consideration that $a=9/32\mathrm{tr}\, R^3$, the definition of $a^{MSY}$ in (\ref{aMSY}) and equations (\ref{norm1}), (\ref{li}), what we have to prove is:
\begin{equation}
\mathrm{tr}\, R^3_{\verb| | |a_i=f_i} = \frac{24}{S}
\end{equation}
where $S$ is the sum of $l_i$, as in the previous Appendix (\ref{esse}).

In this Appendix we will use the notation $b=(x,y)$ to indicate the point $B$ in the plane of $P$ (recall that the Reeb vector can be parametrized as $3(1,x,y)$). With a little abuse of notation, we will call $V_i$ the coordinates $(x_i,y_i)$ in the plane of $P$ of the vertices $V_i$. Hence we have $v_i=V_{i+1}-V_{i}$ and $r_i=V_i-b$. 

To simplify the calculation of $\mathrm{tr}\, R^3$, choose a point $(x^0,y^0)$ in $P$, in general distinct from the ``Reeb point'' $b=(x,y)$. For every field in the quiver gauge theory (in the minimal toric phase described in Section \ref{geometry}) associated with the pair $(i,j)\in C$ consider its R-charge:
\begin{equation}
a_{i,j} \equiv a_{i+1}+a_{i+2} \ldots + a_{j}
\end{equation}  
and perform the substitution $a_i=f_i(x,y)$; we get a rational function of $(x,y)$. Perform the Taylor expansion of this function around the point $(x^0,y^0)$ and denote with $\tilde a_{i,j}$ the truncation of this expansion up to linear terms in $(x,y)$:
\begin{eqnarray}
a_{i,j}(x,y)& = & a_{i,j}(x_0,y_0)+(x_h-x_h^0)\frac{\partial}{\partial x_h}a_{i,j}(x_0,y_0)
                  + O((x_h-x_h^0)^2)\nonumber \\
            & \equiv & \tilde a_{i,j}(x,y) + O((x_h-x_h^0)^2)
\end{eqnarray}
where $x_h$, $h=1,2$, is $x$ or $y$.

We will use the fact that
\begin{equation}
\mathrm{tr}\, R^3=\mathrm{tr}\, R^2 \tilde R 
\label{trunc}
\end{equation}
where $\tilde R$ stands for the vector of truncated R-charges $\tilde a_{i,j}$. In this formula and in the following we always understand the substitutions $a_i=f_i(x,y)$. To prove (\ref{trunc}) note that, by multiplying by $2/S$ equation (\ref{weights}), we get:
\begin{equation}
\sum_{i=1}^d a_i r_i=0, \qquad \Rightarrow \qquad \sum_{i=1}^d a_i V_i=2b
\label{aivi}
\end{equation}  
since $r_i=V_i-b$. Note that this is just equation (2.86) in \cite{MSY}. 
Deriving the last relation with respect to $x$ and/or $y$ we get:
\begin{equation}
\sum_{i=1}^d \left( \frac{\partial}{\partial x_h} \right)^k a_i V_i=0, \qquad \textrm{if} \quad k\geq 2
\end{equation}
where the derivatives can be mixed in $x$, $y$ and have total degree $k\geq 2$. In fact $b=(x,y)$ is linear in $(x,y)$. Deriving instead the relation $\sum_i a_i=2$ we get 
\begin{equation}
\sum_{i=1}^d \left( \frac{\partial}{\partial x_h} \right)^k a_i =0
\end{equation}
The two previous relations tell us that the derivatives of $a_i$ with degree 2 or higher, calculated in any point $(x^0,y^0)$, are baryonic symmetries: see equations (\ref{linear}) and (\ref{sumbaryon}). In the previous Appendix we proved that for any baryonic symmetry $\mathrm{tr}\, R^2 B=0$ for $a_i=f_i(x,y)$. 
Hence we have   
\begin{equation}
\mathrm{tr}\, R^3=\mathrm{tr}\, R^2 \left( \tilde R +  \textrm{higher derivatives} \right)=\mathrm{tr}\, R^2 \tilde R 
\end{equation}
since the other terms in the Taylor expansion are derivatives with degree $k\geq 2$.

In the following we will choose $(x_0,y_0)$ as the first vertex $V_1$ of $P$ and we will calculate $\tilde a_{i,j}(x,y)$ in the point $b=(x,y)$.
So we need to get the explicit expressions for the charges 
\begin{equation}
\tilde a_i (x,y)=a_i(V_1)-r_1 \cdot \vec\nabla a_i(V_1), 
                  \qquad  \vec\nabla a_i=\left( \frac{\partial a_i}{\partial x} , \frac{\partial a_i}{\partial y} \right) 
\label{r1}
\end{equation}
since $r_1=V_1-b=(x_0-x,y_0-y)$ and in the second term we have written the scalar product of this vector with the gradient of $a_i$. The charges of composite fields are obviously given by $\tilde a_{i,j}=\tilde a_{i+1}\ldots +\tilde a_j$.

Let us study first the behavior of $a_i(x,y)=2l_i/S$ when $(x,y)=V_1+tv_1$ approaches the point $V_1$ along the first side of $P$, $0 < t <1$.
Note that $A_1$ goes to zero, whereas the other areas $A_i$ are strictly positive. Hence $l_1$ and $l_2$ goes to $+\infty$ and the other $l_i$ remain finite. 
Hence all $a_i(x,y)=2l_i/S$ different from $a_1$ and $a_2$ goes to zero, since they have a finite numerator and are divided by $S$ which diverges. 
Performing the limit $(x,y)\rightarrow V_1+t v_1$ for $a_1$ and $a_2$ we get:
\begin{eqnarray}
a_1(V_1+tv_1) & = & \frac{2\langle v_d, v_1 \rangle A_2}{A_2\langle v_d,v_1 \rangle + A_d \langle v_1,v_2 \rangle}_{|(x,y)=V_1+t v_1}=2(1-t) \nonumber \\
a_2(V_1+tv_1) & = & \frac{2\langle v_1, v_2 \rangle A_d}{A_2\langle v_d,v_1 \rangle + A_d \langle v_1,v_2 \rangle}_{|(x,y)=V_1+t v_1}=2t
\end{eqnarray} 
where we used $A_d=\langle v_d, t v_1 \rangle$ and $A_2= \langle (1-t)v_1,v_2 \rangle$ when $(x,y)=V_1+t v_1$. Note in particular that for $t=0$, we obtain for the 
vertex $V_1$: $a_1(V_1)=2$ and all other $a_i$ equal to zero. Repeating this analysis on the last side $v_d$ of $P$ we obtain:
\begin{eqnarray}
a_1(V_1-t v_d) & = & 2(1-t) \nonumber \\
a_d(V_1-t v_d) & = & 2t
\end{eqnarray}  
and all other charges $a_i$ equal to zero. 

Deriving the previous relations with respect to $t$ we obtain the gradient of the $a_i$ along the sides $v_1$ and $v_d$ of $P$:
\begin{equation}
\begin{array}{l}
v_1 \cdot \vec\nabla a_1(V_1)=-2 \\
v_d \cdot \vec\nabla a_1(V_1)=2
\end{array}
\quad
\begin{array}{l}
v_1 \cdot \vec\nabla a_2(V_1)=2 \\
v_d \cdot \vec\nabla a_2(V_1)=0
\end{array}
\quad
\begin{array}{l}
v_1 \cdot \vec\nabla a_d(V_1)=0 \\
v_d \cdot \vec\nabla a_d(V_1)=-2
\end{array}
\end{equation}
and zero for all other charges different from $a_d$, $a_1$, $a_2$.
Finally relation (\ref{split}) 
\begin{equation}
r_1=\frac{v_d}{A_d\, l_1}-\frac{v_1}{A_1\, l_1}
\end{equation}
allows to compute $\tilde a_i(x,y)$ from (\ref{r1}):
\begin{equation}
\left\{
\begin{array}{l}
\tilde a_1=2-2\alpha-2\beta\\
\tilde a_d=2\alpha\\
\tilde a_2=2\beta
\end{array}
\right.
\qquad \quad
\left\{
\begin{array}{l}
\alpha \equiv \displaystyle \frac{A_1}{\langle v_d,v_1 \rangle}\\[0.8em]
\beta \equiv \displaystyle \frac{A_d}{\langle v_d,v_1 \rangle}
\end{array}
\right.
\label{aitilde}
\end{equation}
All other $\tilde a_i$ different from $\tilde a_d$, $\tilde a_1$, $\tilde a_2$ are zero.
This fact, together with (\ref{trunc}), allows to disentangle the complex combinatorics and to perform a straightforward, but quite long, computation of $\mathrm{tr}\,R^3$.

So we obtain:
\begin{eqnarray}
& & \hspace{-1em}\mathrm{tr}\, R^3  =  \mathrm{tr} R^2 \tilde R =
                  F + \sum_{(i,j)\in C} \langle v_i,v_j \rangle \left( a_{i,j}-1 \right)^2 \left( \tilde a_{i,j}-1 \right) \nonumber \\
& = & F-\sum_{(i,j)\in C} \langle v_i,v_j \rangle \left( a_{i,j}-1 \right) \left( \tilde a_{i,j}-1 \right)
      +\sum_{h=1}^d a_h \left( \sum_{(i,j)\in C_h} \langle v_i,v_j \rangle \left( a_{i,j}-1 \right) \left( \tilde a_{i,j}-1 \right)\right) \nonumber \\
& \equiv &  F-\sum_{(i,j)\in C} \langle v_i,v_j \rangle \left( a_{i,j}-1 \right) \left( \tilde a_{i,j}-1 \right)
            +\sum_{h=1}^d a_h c_h  
\label{step1}
\end{eqnarray}
where $c_h$ are defined by the last equality.
With similar tricks as in previous Appendices, we see that:
\begin{equation}
c_{j+1}-c_{j}=\langle v_j, T_j \rangle
\end{equation}
where the vector $T_j$ is:
\begin{eqnarray}
T_j & = & \bar T_j+v_j-\frac{4}{S}\frac{v_j}{A_j} \nonumber \\
\bar T_j & = & v_{j+1} \left( a_{j+1}-1 \right) \left( \tilde a_{j+1}-1 \right)\ldots 
         + v_{k_j}\left( a_{j+1}\ldots +a_{k_j}-1 \right)\left( \tilde a_{j+1}\ldots + \tilde a_{k_j}-1 \right) \nonumber \\
    &   & +  v_{j-1} \left( a_j -1 \right) \left( \tilde a_j -1 \right) \ldots +v_{k_j+1}\left( a_j \ldots+ a_{k_j+2} -1 \right)
         \left(\tilde a_j \ldots + \tilde a_{k_j+2} -1 \right) \nonumber \\
    & = & v_{j+1} \left( a_{j+1}-1 \right) \left( \tilde a_{j+1}-1 \right)+v_{j+2} \left( a_{j+1}+a_{j+2}-1 \right)               \left( \tilde a_{j+1}+\tilde a_{j+2}-1 \right)\ldots \nonumber \\
    &   & + v_{j-1} \left( a_{j+1}+a_{j+2} \ldots +a_{j-1} -1 \right) 
          \left( \tilde a_{j+1}+ \tilde a_{j+2} \ldots + \tilde a_{j-1} -1 \right)
\end{eqnarray}
and the pieces proportional to $v_j$ have been introduced for later convenience. For the difference of consecutive $\bar T_j$ we obtain:
\begin{eqnarray}
 & & \hspace{-1.5em}\bar T_2 -\bar T_1 = \nonumber \\[0.2em]
 & = & v_3 \left( a_3 -1 \right) \left( \tilde a_3 -1 \right)+ v_4\left( a_3+a_4 -1 \right) \left( \tilde a_3+\tilde a_4 -1 \right) \nonumber \\
 &   &   \ldots + v_1\left( a_3+a_4\ldots+a_1 -1 \right) \left( \tilde a_3+\tilde a_4\ldots+ \tilde a_1 -1 \right) \nonumber \\
 &  & -v_2 \left( a_2-1 \right)\left( \tilde a_2-1 \right)-v_3 \left( a_2+a_3 -1 \right) \left( \tilde a_2+\tilde a_3 -1 \right) \nonumber \\
 &  &    \ldots -v_d \left( a_2+a_3\ldots+a_d -1 \right) \left( \tilde a_2+\tilde a_3\ldots+ \tilde a_d -1 \right) \nonumber \\[0.2em] 
 & = & - a_2 \tilde a_2 \left[v_2+v_3 \ldots +v_d \right]+v_2 a_2 + v_2 \tilde a_2 - v_2 + v_1 \left(a_2-1 \right) \left(\tilde a_2-1 \right) \nonumber \\
 &   & -a_2 \left[ v_3 \left( \tilde a_3 -1 \right)+ v_4 \left( \tilde a_3 + \tilde a_4 -1 \right) \ldots
        + v_d \left( \tilde a_3 + \tilde a_4\ldots + \tilde a_d -1 \right) \right] \nonumber \\
 &   & -\tilde a_2 \left[ v_3 \left( a_3 -1 \right)+ v_4 \left( a_3 +a_4 -1 \right) \ldots + v_d \left( a_3 + a_4\ldots +a_d -1 \right) \right] \nonumber \\[0.2em]
 & = & -a_2 \left[ v_3 \left( \tilde a_3 -1 \right)+ v_4 \left( \tilde a_3 + \tilde a_4 -1 \right) \ldots
        + v_1 \left( \tilde a_3 + \tilde a_4\ldots + \tilde a_1 -1 \right) \right] \nonumber \\
 &   & -\tilde a_2 \left[ v_3 \left( a_3 -1 \right)+ v_4 \left( a_3 +a_4 -1 \right) \ldots + v_1 \left( a_3 + a_4\ldots +a_1 -1 \right) \right] \nonumber \\
 &   & +v_2 a_2 +v_2 \tilde a_2 + v_1-v_2 \nonumber \\[0.2em]
 & = & -a_2 \left[ \tilde a_1 v_1+ \tilde a_d \left( v_1+v_d \right) \ldots + \tilde a_3 \left( v_1+v_d \ldots + v_3  \right) \right] \nonumber \\
 &   & -\tilde a_2 \left[ a_1 v_1+a_d \left( v_1+v_d \right) \ldots + a_3 \left( v_1+v_d \ldots + v_3  \right) \right] + v_1- v_2 \nonumber \\[0.2em]
 & =  & v_1-v_2+ a_2 \left( \sum_{i=1}^d \tilde a_i (V_i-V_2) \right)+ \tilde a_2 \left( \sum_{i=1}^d a_i (V_i-V_2) \right) \nonumber \\[0.2em]
 & = & v_1-v_2 - 2 a_2 r_2 - 2 \tilde a_2 r_2 =  v_1-v_2 - \frac{4}{S}\frac{v_1}{A_1}+ \frac{4}{S}\frac{v_2}{A_2} - 2 \tilde a_2 r_2
\end{eqnarray}
where in the last step we used (\ref{aivi}) (which is also true for $\tilde a_i$, as one deduces from its Taylor expansion up to linear terms), $r_2=V_2-b$ and (\ref{split}).
By relabelling indices:
\begin{equation}
T_{j+1}-T_j=-2 \tilde a_{j+1} r_{j+1}
\end{equation}
Note that
\begin{equation}
T_2=T_3 \ldots = T_{d-1} \, = \, T_1 -2 \tilde a_2 r_2
\end{equation}
since $\tilde a_i$ are zero for $i=3,4, \ldots d-1$. 
We obtain then
\begin{eqnarray}
 &   & \hspace{-1.5em} \sum_{h=1}^d a_h c_h = c_1 \left( a_1+a_2 \ldots +a_d \right)+ a_2 \left( c_2-c_1 \right) \ldots + a_d \left( c_d-c_1 \right) \nonumber \\[0.2em]
 & = & 2 c_1 + a_2 \langle v_1, T_1 \rangle + a_3 \left( \langle v_1, T_1 \rangle + \langle v_2, T_2 \rangle \right) 
       \ldots + a_d \left( \langle v_1, T_1 \rangle \ldots  + \langle v_{d-1}, T_{d-1} \rangle \right) \nonumber \\[0.2em]
 & = & 2 c_1 +\langle a_2 v_1 +a_3 \left( v_1+v_2 \right) \ldots + a_d \left( v_1+v_2 \ldots v_d \right), T_1 \rangle \nonumber \\
 &   & -2 \tilde a_2 \langle a_3 v_2+ a_4 \left( v_2+v_3 \right)\ldots +a_d \left( v_2+v_3 \ldots + v_{d-1} \right) ,r_2 \rangle \nonumber \\[0.2em]
 & = & 2 c_1 + \langle \sum_{i=1}^d a_i \left( V_i-V_1 \right) , T_1 \rangle -2 \tilde a_2 \langle \sum_{i=1}^d a_i \left( V_i-V_2 \right) , r_2 \rangle
       + 2 \tilde a_2 \langle \left( V_1-V_2 \right) a_1 ,r_2 \rangle \nonumber \\[0.2em]
 & = & 2 c_1 -2 \langle r_1, T_1 \rangle + 4 \tilde a_2 \langle r_2, r_2 \rangle -2 \tilde a_2 a_1 \langle v_1, r_2 \rangle \nonumber \\[0.2em]
 & = & 2 c_1 -2 \langle r_1, T_1 \rangle + 4 a_1 \frac{A_d A_1}{\langle v_d, v_1 \rangle}= 2 c_1 -2 \langle r_1, T_1 \rangle + \frac{4 a_1}{l_1} \nonumber \\[0.2em]
 & = &  2 c_1 -2 \langle r_1, T_1 \rangle + \frac{8}{S}
\label{ahch1}
\end{eqnarray}
where we have used the explicit expression (\ref{aitilde}) for $\tilde a_2$, and performed the substitution $a_1=2 l_1/S$.

From the definition we now compute:
\begin{eqnarray}
 &   & \hspace{-1.5em} \bar T_1 = \nonumber \\[0.2em]
 & = &  v_2 \left( a_2-1 \right)\left(\tilde a_2-1 \right)+v_3 \left( a_2+a_3 -1 \right)\left(\tilde a_2+ \tilde a_3 -1 \right) \ldots \nonumber \\
 &   &  + v_d \left( a_2+a_3\ldots +a_d -1 \right)\left(\tilde a_2 +\tilde a_3 \ldots +\tilde a_d -1 \right) \nonumber \\[0.2em]
 & = & \left( \tilde a_2-1 \right)\left[v_2 \left( a_2-1 \right) + v_3 \left( a_2+a_3-1 \right)\ldots + v_d\left( a_2 \ldots + a_d -1 \right) \right] 
       + \tilde a_d v_d (1-a_1) \nonumber \\[0.2em]
 & = & \left( \tilde a_2-1 \right)\left[ -\left( v_2+v_3 \ldots + v_d \right)-\sum_{i=1}^d a_i \left(V_i-V_1 \right) \right] + \tilde a_d v_d (1-a_1) \nonumber \\[0.2em]
 & = & \left( \tilde a_2-1 \right) \left( v_1 + 2 r_1 \right) + \tilde a_d v_d (1-a_1) 
\end{eqnarray}
and hence 
\begin{eqnarray}
\langle r_1, T_1 \rangle & = & \langle r_1,\left( \tilde a_2-1 \right) \left( v_1 + 2 r_1 \right)
                                + \tilde a_d v_d \left( 1-a_1 \right) + v_1 -\frac{4}{S}\frac{v_1}{A_1} \rangle \nonumber \\[0.2em]
& = & \tilde a_2 \langle r_1,v_1 \rangle + \tilde a_d \left( 1-a_1 \right)\langle r_1,v_d \rangle -\frac{4}{S\,A_1} \langle r_1,v_1 \rangle \nonumber \\[0.2em]
& = & \frac{2 A_d A_1}{\langle v_d,v_1 \rangle}+ \frac{2 A_d A_1}{\langle v_d,v_1 \rangle} (1-a_1)-\frac{4}{S}
      = \frac{2}{l_1} \left(2-\frac{2 l_1}{S} \right)-\frac{4}{S} \nonumber \\[0.2em]
& = & \frac{4}{l_1}-\frac{8}{S}
\end{eqnarray}
where again we have used the explicit expressions for $\tilde a_i$ in (\ref{aitilde}) and the substitution $a_1=2l_1/S$. 
Collecting pieces together we obtain for (\ref{ahch1}):
\begin{equation}
\sum_{h=1}^d a_h c_h = 2 c_1+\frac{24}{S}-\frac{8}{l_1}
\label{ahch}
\end{equation}

Going back to (\ref{step1}) we obtain:
\begin{eqnarray}
 &   & \hspace{-1.5em}\mathrm{tr}\, R^3  = \nonumber \\
 & = &  F-\sum_{(i,j)\in C} \langle v_i,v_j \rangle \left( a_{i,j}-1 \right) \left( \tilde a_{i,j}-1 \right)
        +\sum_{h=1}^d a_h c_h \nonumber \\
 & = &  F + \sum_{(i,j)\in C} \langle v_i,v_j \rangle \left( a_{i,j}-1 \right) - \sum_{(i,j)\in C} \langle v_i,v_j \rangle \left( a_{i,j}-1 \right)\tilde a_{i,j}
        +\sum_{h=1}^d a_h c_h \nonumber \\
 & = &  - \sum_{(i,j)\in C} \langle v_i,v_j \rangle \left( a_{i,j}-1 \right)\tilde a_{i,j}+\sum_{h=1}^d a_h c_h
\label{step2} 
\end{eqnarray}
where in the last step we used $\mathrm{tr}\, R=0$ (\ref{tracer}).

Let us expand the first term in the previous equality; we use the explicit form (\ref{aitilde}) of the $\tilde a_{i}$ putting in evidence the factors of $2$, $\alpha$ and $\beta$:
\begin{eqnarray}
 &   & \hspace{-1.5em}\sum_{(i,j)\in C} \langle v_i,v_j \rangle \left( a_{i,j}-1 \right)\tilde a_{i,j} = 
       \nonumber \\[0.2em]
 & = & 2 \sum_{(i,j)\in C_1} \langle v_i,v_j \rangle \left( a_{i,j}-1 \right)\nonumber \\
 &   & + 2 \alpha \left( \sum_{(i,j)\in \left(C_d-C_1\right)} \langle v_i,v_j \rangle \left( a_{i,j}-1 \right) \right)
       - 2 \alpha \left( \sum_{(i,j)\in \left(C_1-C_d\right)} \langle v_i,v_j \rangle \left( a_{i,j}-1 \right) \right)         \nonumber \\
 &   & + 2 \beta \left( \sum_{(i,j)\in \left(C_2-C_1\right)} \langle v_i,v_j \rangle \left( a_{i,j}-1 \right) \right) 
       - 2 \beta \left( \sum_{(i,j)\in \left(C_1-C_2\right)} \langle v_i,v_j \rangle \left( a_{i,j}-1 \right) \right)   
       \nonumber \\
 &    &  \label{term1}    
\end{eqnarray}
Expanding the factor $2 c_1$ in (\ref{ahch}) we obtain:
\begin{eqnarray}
 &   & \hspace{-1.5em} 2c_1 = \nonumber \\[0.2em]
 & = & 2 \sum_{(i,j)\in C_1} \langle v_i,v_j \rangle \left( a_{i,j}-1 \right)\left(\tilde a_{i,j}-1 \right) 
       \nonumber \\[0.2em]
 & = & - 2 \sum_{(i,j)\in C_1} \langle v_i,v_j \rangle \left( a_{i,j}-1 \right)
       + 2 \sum_{(i,j)\in C_1} \langle v_i,v_j \rangle \left( a_{i,j}-1 \right) \tilde a_{i,j}  \nonumber \\[0.2em]
 & = & - 2 \sum_{(i,j)\in C_1} \langle v_i,v_j \rangle \left( a_{i,j}-1 \right)
       + 4 \sum_{(i,j)\in C_1} \langle v_i,v_j \rangle \left( a_{i,j}-1 \right) \nonumber \\
 &   & - 4 \alpha \left( \sum_{(i,j)\in \left(C_1-C_d\right)} \langle v_i,v_j \rangle \left( a_{i,j}-1 \right) \right)
       - 4 \beta \left( \sum_{(i,j)\in \left(C_1-C_2\right)} \langle v_i,v_j \rangle \left( a_{i,j}-1 \right) \right)
       \nonumber \\
 &   & \label{term2}         
\end{eqnarray}
Then equation (\ref{step2}), using (\ref{ahch}), (\ref{term1}), (\ref{term2}), becomes:
\begin{eqnarray}
 &   &\hspace{-1.5em}\mathrm{tr}\, R^3 = \nonumber \\ 
 & = &  -2 \alpha  \left(  \sum_{(i,j)\in \left(C_1-C_d\right)} + 
       \sum_{(i,j)\in \left(C_d-C_1\right)} \right)\langle v_i,v_j \rangle \left( a_{i,j}-1 \right) \nonumber \\ 
 &   &  -2 \beta   \left(  \sum_{(i,j)\in \left(C_1-C_2\right)} + 
       \sum_{(i,j)\in \left(C_2-C_1\right)} \right)\langle v_i,v_j \rangle \left( a_{i,j}-1 \right)                            +\frac{24}{S}-\frac{8}{l_1} 
\label{step3}       
\end{eqnarray} 
and it is easy to compute:
\begin{eqnarray}
 &   & \hspace{-1.5em} \left(  \sum_{(i,j)\in \left(C_1-C_d\right)} + 
       \sum_{(i,j)\in \left(C_d-C_1\right)} \right)\langle v_i,v_j \rangle \left( a_{i,j}-1 \right) = 
       \nonumber \\[0.2em]
 & = & \langle v_d,v_1 \rangle \left( a_1-1 \right) + \langle v_d,v_2 \rangle \left( a_1+a_2-1 \right) \ldots
       + \langle v_d,v_{k_d} \rangle \left( a_1+a_2\ldots +a_{k_d}-1 \right) \nonumber \\
 &   & - \langle v_d,v_{d-1} \rangle \left( a_d-1 \right) 
       %- \langle v_d,v_{d-2} \rangle \left( a_d + a_{d-1}-1 \right) 
      \ldots - \langle v_d,v_{k_d+1} \rangle \left( a_d + a_{d-1}\ldots+a_{k_d+2}-1 \right) \nonumber \\[0.2em]
 & = & \langle v_d, v_1 \left( a_1-1 \right)+  v_2 \left( a_1+a_2-1 \right)\ldots 
      + v_{d-1} \left( a_1+a_2\ldots +a_{d-1}-1 \right)\rangle \nonumber \\[0.2em]
 & = & \langle v_d, -\left( v_1+v_2 \ldots +v_{d-1} \right) + a_1 \left(v_1+v_2\ldots +v_{d-1}\right)  
       + a_2 \left(v_2\ldots +v_{d-1}\right) \ldots +a_{d-1}v_{d-1} \nonumber \\[0.2em]  
 & = & \langle v_d, -\sum_{i=1}^d a_i \left( V_i - V_d \right) \rangle = 2 \langle v_d, r_d \rangle =-2 A_d             \end{eqnarray}
and similarly, with the opportune changes in the indexes:
\begin{equation}
\left(  \sum_{(i,j)\in \left(C_2-C_1\right)} + 
       \sum_{(i,j)\in \left(C_1-C_2\right)} \right)\langle v_i,v_j \rangle \left( a_{i,j}-1 \right) = 
       -2 A_1
\end{equation}   
Finally equation (\ref{step3}) becomes:
\begin{eqnarray}
\mathrm{tr}\, R^3 & = & \frac{24}{S}-\frac{8}{l_1}-2\frac{A_1}{\langle v_d,v_1 \rangle}\left(-2A_d \right) 
                        -2\frac{A_d}{\langle v_d,v_1 \rangle}\left(-2A_1 \right)  \nonumber \\
& = & \frac{24}{S}
\end{eqnarray}
and this concludes our proof.

\end{document}